\documentclass[a4paper,UKenglish,thm-restate,cleveref]{lipics-v2021}
\pdfoutput=1

\title{Correct Cryptocurrency ASIC Pricing: Are Miners Overpaying?}

\author{Aviv Yaish}{The Hebrew University, Israel}{aviv.yaish@mail.huji.ac.il}{https://orcid.org/0000-0002-7971-2494}{}
\author{Aviv Zohar}{The Hebrew University, Israel}{avivz@cs.huji.ac.il}{https://orcid.org/0000-0001-8539-9222}{}

\authorrunning{A. Yaish and A. Zohar} 

\Copyright{Aviv Yaish and Aviv Zohar}


\category{}
\relatedversion{See the full version of the paper for proofs and additional details.}
\relatedversiondetails[cite=yaish2023pricing]{Full Version}{https://doi.org/10.48550/arXiv.2002.11064}

\acknowledgements{This research was supported by the Ministry of Science \& Technology, Israel.}

\nolinenumbers 

\EventEditors{Joseph Bonneau and S. Matthew Weinberg}
\EventNoEds{2}
\EventLongTitle{5th Conference on Advances in Financial Technologies (AFT 2023)}
\EventShortTitle{AFT 2023}
\EventAcronym{AFT}
\EventYear{2023}
\EventDate{October 23-25, 2023}
\EventLocation{Princeton, NJ, USA}
\EventLogo{}
\SeriesVolume{282}
\ArticleNo{2}

\newif\ifcomments
\newif\ifanonymous
\newif\iflong

\commentsfalse
\anonymousfalse
\longtrue

\usepackage{mathtools}      
\usepackage[ruled]{algorithm2e} 
\usepackage{forest}         
\usepackage{graphicx}       
\graphicspath{ {./images/} }
\usepackage{amsthm}         
\usepackage{thmtools}   
\usepackage{xcolor}         
\hypersetup{                
    colorlinks,
    linkcolor={red!50!black},
    citecolor={blue!50!black},
    urlcolor={blue!80!black}
}

\usepackage[colorinlistoftodos,prependcaption,textsize=tiny,textwidth=\marginparwidth]{todonotes}

\crefname{definition}{Definition}{Definitions}
\crefname{theorem}{Theorem}{Theorems}
\crefname{claim}{Claim}{Claims}
\crefname{lemma}{Lemma}{Lemmas}
\crefname{corollary}{Corollary}{Corollaries}
\crefname{example}{Example}{Examples}
\crefname{remark}{Remark}{Remarks}
\crefname{glossary}{Glossary}{Glossaries}

\DeclarePairedDelimiter{\ceil}{\lceil}{\rceil}

\newcommand{\define}{\stackrel{\mathclap{\tiny\mbox{def}}}{=}}

\usepackage[symbols,acronym,nonumberlist,nogroupskip,stylemods={mcols,longbooktabs}]{glossaries-extra}
\makenoidxglossaries

\setabbreviationstyle[acronym]{long-short}  
\glssetcategoryattribute{acronym}{nohyper}{true} 
\newacronym{DeFi}{DeFi}{decentralized finance}
\newacronym{PoW}{PoW}{Proof-of-Work}
\newacronym{PoS}{PoS}{Proof-of-Stake}
\newacronym{MEV}{MEV}{miner-extractable value}
\newacronym{block-DAG}{block-DAG}{block directed-acyclic-graph}
\newacronym{DAA}{DAA}{difficulty-adjustment algorithm}
\newacronym{MDP}{MDP}{Markov decision-process}
\newacronym{DQL}{DQL}{Deep-Q-learning}
\newacronym{RL}{RL}{reinforcement learning}
\newacronym{ML}{ML}{machine learning}
\newacronym{AI}{AI}{artificial intelligence}
\newacronym{PDF}{PDF}{probability density function}
\newacronym{CDF}{CDF}{cumulative density function}
\newacronym{ASIC}{ASIC}{application specific integrated circuit}
\newacronym{NIST}{NIST}{National Institute of Standards and Technology}
\newacronym{BTC}{BTC}{Bitcoin}
\newacronym{USD}{USD}{United States Dollar}
\newacronym{kWh}{kWh}{Kilowatt-hour}
\newacronym{NPV}{NPV}{net present value}

\glsxtrnewsymbol[description={
    The symbol for the Bitcoin cryptocurrency.
}]{bitcoin}{
    \ensuremath{%
        \leavevmode
	    \vtop{\offinterlineskip 
		\setbox0=\hbox{B}%
		\setbox2=\hbox to\wd0{\hfil\hskip-.03em
			\vrule height .3ex width .15ex\hskip .08em
			\vrule height .3ex width .15ex\hfil}
		\vbox{\copy2\box0}\box2}%
    }
}
\newcommand{\bitcoin}{{\gls[hyper=false]{bitcoin}}}

\glsxtrnewsymbol[description={
    The symbol for the Ethereum cryptocurrency.
}]{ethereum}{
    \ensuremath{\Xi}
}

\glsxtrnewsymbol[description={
    The difficulty of mining a block.
}]{difficulty}{
    \ensuremath{D}
}

\glsxtrnewsymbol[description={
    The reward received for mining a block, in tokens.
}]{blockreward}{
    \ensuremath{R}
}
\newcommand{\blockreward}[1]{{\gls[hyper=false]{blockreward}_{#1}}}

\glsxtrnewsymbol[description={
    The yearly risk-free rate.
}]{riskfreerate}{
    \ensuremath{r}
}
\newcommand{\riskfreerate}{{\gls[hyper=false]{riskfreerate}}}

\glsxtrnewsymbol[description={
    The value of a single coin, in USD.
}]{coinprice}{
    \ensuremath{c}
}
\newcommand{\coinprice}[1]{{\gls[hyper=false]{coinprice}_{#1}}}

\glsxtrnewsymbol[description={
    Price of electricity, in US Dollars per kilowatt-hour.
}]{electricityprice}{
    \ensuremath{e}
}
\newcommand{\electricityprice}[1]{{\gls[hyper=false]{electricityprice}_{#1}}}

\glsxtrnewsymbol[description={
    Kilowatt-hours required for the computation of a single mining opportunity.
}]{efficiency}{
    \ensuremath{\varphi}
}
\newcommand{\efficiency}{{\gls[hyper=false]{efficiency}}}

\glsxtrnewsymbol[description={
    The probability of an increase in the coin's price.
}]{upprobability}{
    \ensuremath{q}
}
\newcommand{\upprobability}{{\gls[hyper=false]{upprobability}}}

\glsxtrnewsymbol[description={
    The probability of a decrease in the coin's price.
}]{downprobability}{
    \ensuremath{1-q}
}
\newcommand{\downprobability}{{\gls[hyper=false]{downprobability}}}

\glsxtrnewsymbol[description={
    The multiplicative factor by which the coin's price can increase.
}]{upcoinprice}{
    \ensuremath{\Delta}
}
\newcommand{\upcoinprice}{{\gls[hyper=false]{upcoinprice}}}

\glsxtrnewsymbol[description={
    The multiplicative factor by which the coin's price can decrease.
}]{downcoinprice}{
    \ensuremath{\delta}
}
\newcommand{\downcoinprice}{{\gls[hyper=false]{downcoinprice}}}

\glsxtrnewsymbol[description={
    The ASIC's mortality distribution.
}]{mortality}{
    \ensuremath{M}
}
\newcommand{\mortality}{{\gls[hyper=false]{mortality}}}

\glsxtrnewsymbol[description={
    Government issued bonds, yielding the risk-free rate.
}]{bonds}{
    \ensuremath{B}
}
\newcommand{\bonds}[1]{{\gls[hyper=false]{bonds}_{#1}}}

\newcommand{\coinamount}{a}

\glsxtrnewsymbol[description={
    Amount of coins to hold a short position on.
}]{shortcoins}{
    \ensuremath{\coinamount}
}
\newcommand{\shortcoins}[1]{{\gls[hyper=false]{shortcoins}_{#1}}}

\glsxtrnewsymbol[description={
    Amount of coins to hold a long position on.
}]{longcoins}{
    \ensuremath{\overline{\coinamount}}
}
\newcommand{\longcoins}[1]{{\gls[hyper=false]{longcoins}}_{#1}}

\glsxtrnewsymbol[description={
    The hash-rate of the ASIC to price, in hashes-per-second.
}]{hashrate}{
    \ensuremath{h}
}
\newcommand{\hashrate}{{\gls[hyper=false]{hashrate}}}

\glsxtrnewsymbol[description={
    The global hash-rate active on the network, in hashes-per-second.
}]{globalhashrate}{
    \ensuremath{H}
}
\newcommand{\globalhashrate}[1]{{\gls[hyper=false]{globalhashrate}}\left({#1}\right)}

\newcommand{\hashratio}[1]{\frac{\hashrate}{\globalhashrate{#1} + \hashrate}}

\glsxtrnewsymbol[description={
    Value of a portfolio, in US dollars.
}]{portfoliovalue}{
    \ensuremath{\Phi}
}
\newcommand{\portfoliovalue}[1]{{\gls[hyper=false]{portfoliovalue}}\left({#1}\right)}

\glsxtrnewsymbol[description={
    Multiplicative return of a portfolio.
}]{portfolioreturn}{
    \ensuremath{\rho}
}
\newcommand{\portfolioreturn}[1]{{\gls[hyper=false]{portfolioreturn}}\left({#1}\right)}

\newcommand{\valueletter}{V}

\glsxtrnewsymbol[description={
    Value of a mining opportunity, in US dollars.
}]{optionvalue}{
    \ensuremath{\valueletter}
}
\newcommand{\optionvalue}[1]{{\gls[hyper=false]{optionvalue}}\left({#1}\right)}

\glsxtrnewsymbol[description={
    Value of an ASIC, in US dollars.
}]{asicvalue}{
    \ensuremath{\valueletter_{ASIC}}
}
\newcommand{\asicvalue}[1]{{\gls[hyper=false]{asicvalue}}\left({#1}\right)}

\newcommand{\imitatingportfoliovalue}[1]{\overline{\Phi}\left({#1}\right)}

\newcommand{\optioncoefficient}{\gamma}
\newcommand{\downcoefficient}{\optioncoefficient_\downarrow}
\newcommand{\upcoefficient}{\optioncoefficient_\uparrow}
    
\newcommand{\turnReg}{\tau}
\newcommand{\firstturn}{k}
\newcommand{\finalturn}{t}
\newcommand{\asicturn}{s}

\begin{document}
\maketitle
\begin{abstract}
	Cryptocurrencies that are based on \gls{PoW} often rely on special purpose hardware to perform so-called \emph{mining} operations that secure the system, with miners receiving freshly minted tokens as a reward for their work. A notable example of such a cryptocurrency is Bitcoin, which is primarily mined using \gls{ASIC} based machines. Due to the supposed profitability of cryptocurrency mining, such hardware has been in great demand in recent years, in-spite of high associated costs like electricity.
	
	In this work, we show that because mining rewards are given in the mined cryptocurrency, while expenses are usually paid in some fiat currency such as the \gls{USD}, cryptocurrency mining is in fact a bundle of \emph{financial options}. When exercised, each option converts electricity to tokens.
	
	We provide a method of pricing mining hardware based on this insight, and prove that any other price creates arbitrage. Our method shows that contrary to the popular belief that mining hardware is worth less if the cryptocurrency is highly volatile, the opposite effect is true: volatility \emph{increases} value. Thus, if a coin's volatility decreases, some miners may leave, affecting security.
	
	We compare the prices produced by our method to prices obtained from popular tools currently used by miners and show that the latter only consider the expected returns from mining, while neglecting to account for the inherent risk in mining, which is due to the high exchange-rate volatility of cryptocurrencies.
	
	Finally, we show that the returns made from mining can be imitated by trading in bonds and coins, and create such imitating investment portfolios. Historically, realized revenues of these portfolios have \emph{outperformed} mining, showing that indeed hardware is mispriced.
\end{abstract}

\section{Introduction}
\label{sec:Introduction}
The cryptocurrency boom was heralded by the arrival of Bitcoin \cite{nakamoto2008bitcoin}, which introduced the idea of a decentralized currency to the mainstream. Bitcoin relies on pseudonymous users called \emph{miners} to operate the cryptocurrency's ledger in a decentralized manner. In particular, a computationally-heavy mechanism called \gls{PoW} is used to achieve consensus between miners on the system's state and secure it from various attacks \cite{sompolinsky2016bitcoins}.

Miners are rewarded for their work via a form of computation based lottery, yielding additional rewards the more they compute on behalf of the system. These mining rewards have led to an arms-race in which miners purchase increasingly efficient and performant hardware \cite{bedfordtaylor2017evolution}. Today's Bitcoin mining is mostly performed in industrial-scale mining ``farms'' hosting \glspl{ASIC} tailor-made for mining \cite{rauchs2017global}.

To stay competitive, miners buy mining rigs in advance at a significant capital expenditure, and they go to great lengths to keep their hardware's electricity cost at a minimum \cite{rauchs2017global}. Thus, miners turn off their machines if it is unprofitable to mine \cite{news2021bitcoin,sigalos2022as,helman2020how,harper2022hashrate}, and even transport hardware between remote locations on a seasonal basis to save on electricity \cite{redman2019chinese,chipolina2020bitcoins,haig2020btc}.

\begin{figure}
    \centering
	\includegraphics[trim={0.0cm 0.0cm 0.0cm 0.0cm},clip,width=0.7\linewidth]{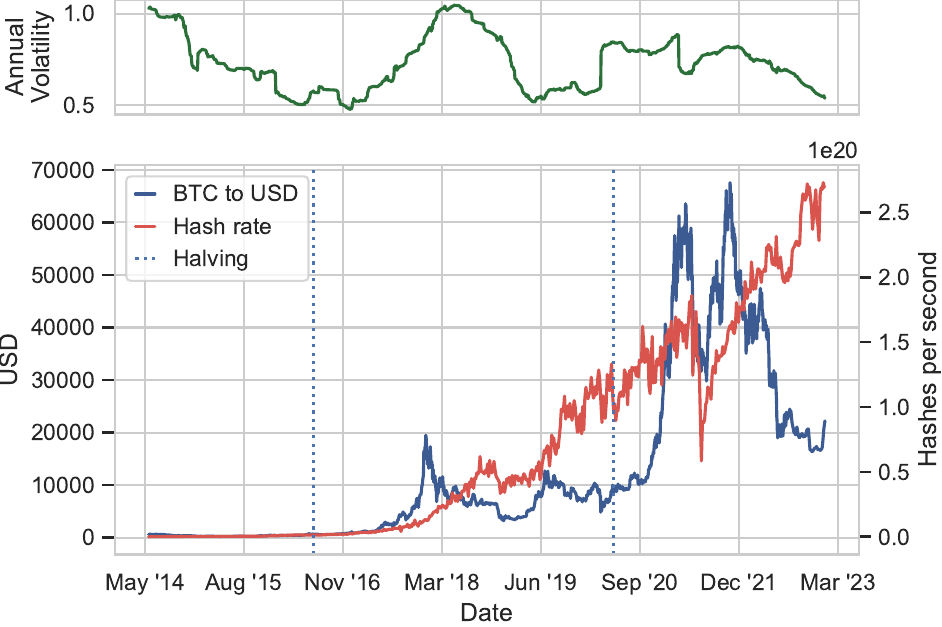}
	\caption{Bitcoin's exchange rate, annual volatility and global hash-rate, as functions of time.}
	\label{fig:BtcToUsdAndHash}
\end{figure}

The profits derived from mining are highly volatile as they depend on the erratic exchange-rate of the cryptocurrency received as reward (see \cref{fig:BtcToUsdAndHash}) and on the level of competition from other miners. These factors make mining a risky investment and may indirectly hurt the cryptocurrency if fewer miners are there to secure it.

Despite the high-risk returns from mining, mining calculators utilize basic techniques to evaluate the price of mining hardware. These na\"ive approaches revolve around a metric called the \emph{hashprice}, which assumes that the currency's exchange-rate is \emph{constant} and ignore the associated risk.

\begin{definition}[Hashprice]
    \label{definition:hashprice}
    The \emph{hashprice} of a specific mining machine is the expected profit that it produces per unit of computation, given that the exchange-rate of the mined token and the computational power mining it are constant until the end-of-life of the machine.
\end{definition}

This metric was introduced by the Luxur mining pool \cite{index2022bitcoin}, and is widely used by the community \cite{mining2021what,sherwood2022why,company2022mining,protocol2022what}. Indeed, this metric is used by the top 8 websites which correspond to the keywords \emph{mining calculator} \cite{value2022miners,index2022profitability,minerstat2022mining,whattomine2022crypto,cryptocompare2022bitcoin,nicehash2022profitability,coinwarz2022bitcoin,company2022bitcoin,company2022bitcoin}, according to the web traffic analyzer \emph{similarweb} \cite{similarweb2022website}. These sites were frequently recommended on mining-related resources \cite{zucchi2022is,taserz2019asic.to}.

\subsection{Our Approach}
In contrast to using the expected rewards as captured by the hashprice, we advance a more nuanced approach for evaluating cryptocurrency mining hardware, such as \glspl{ASIC} that accounts for risk attitudes regarding the exchange rate. Risk attitudes are subjective, and hard to measure but are reflected in the exchange rate itself. We utilize tools from financial option pricing to incorporate the market's risk attitude into the ASIC pricing model. 

Specifically, we show that using publicly available information about the market (such as the interest-rate) and information about the efficiency of hardware (such as hash rate and power consumption of each device), one can derive a \emph{correct} price for the hardware. This price is correct in the sense that any other price creates \emph{arbitrage} and thus allows market forces to earn a risk-free profit.

Then, we construct an investment portfolio of tokens and bonds which \emph{imitates} an \gls{ASIC} and provides the same profits as a given mining machine. This portfolio has the same price as the correct price of the machine which it imitates. We empirically evaluate such imitating portfolios over historical data and show that they earn more than the corresponding \glspl{ASIC}, while costing less.

To obtain the correct price and an imitating portfolio which has an equal cost, we prove \glspl{ASIC} are equivalent to a bundle of \emph{financial options} that allow their owners to exchange electricity for coins at different points in time. Then, we present algorithms which compute the correct hardware cost and the corresponding imitating portfolio.

\paragraph*{Summary of Contributions}
\begin{itemize}
    \item \textbf{An economic model for \gls{PoW} cryptocurrency mining.} We model mining while accounting for the inherent risk due to the volatile exchange-rate of cryptocurrencies. At first glance it may seem that higher volatility in rewards implies higher risk for miners, which may devalue mining machines, but in fact, we show that mining machines increase in value if the cryptocurrency is more volatile. This is because if the exchange rate plummets, the losses of miners are bounded (they can always shut off their machines and avoid paying for electricity), but if exchange rates increase steeply their gains can be significant.
    
    \item \textbf{An algorithm for pricing cryptocurrency mining hardware.} Using our model, we provide an algorithm that computes the price of an \gls{ASIC} given its specifications and market parameters, without relying on subjective measures like a miner's risk preference or the expected exchange-rate of the mined cryptocurrency. We prove that any other price creates arbitrage.
    
    \item \textbf{The effects of risk and delay on the price of mining hardware.} We quantify the impact of the volatility of Bitcoin's exchange-rate and the delays miners frequently face when ordering new hardware on the value of mining machines.
    
    \item \textbf{Imitating portfolios for mining hardware.} We construct an \emph{imitating portfolio} consisting of coins and bonds, which ideally provides the same returns as a specific \gls{ASIC}.
    
    \item \textbf{Empirical evaluation.} We make a three-way comparison between historical hardware prices, the correct prices obtained by our results, and the costs of the corresponding imitating portfolios. Historically, our portfolios earn \emph{more} than \glspl{ASIC} while costing less, even when considering trading fees. These results imply that \glspl{ASIC} are overpriced.
\end{itemize}

\subsection{Motivating Example}
To motivate our work, we show in \cref{exm:SimpleExample} that the hashprice metric (given in \cref{definition:hashprice}) is flawed. Furthermore, we show that more complex hardware pricing methods such as using the expected profit of a mining machine produce incorrect prices that create arbitrage opportunities.

\begin{example}
    \label{exm:SimpleExample}
	A vendor offers the option of using its \gls{ASIC} tomorrow to mine a single block. The vendor assures that if the \gls{ASIC} is turned on, it will earn exactly 1 Bitcoin (henceforth denoted as \glsxtrshort{BTC} or $\bitcoin$), and will require $\$250$ worth of electricity. For simplicity, let the interest rate be $1$.
	
	Assume bitcoin's value starts at $\$400$ today, and will either double or halve tomorrow with equal probability. Note that per the definition of the hashrate metric (see \cref{definition:hashprice}), the price of the option is $\$400 - \$250 = \$150$. Furthermore, this price does not depend on the exchange-rate's random walk, but rather solely on its current value.

    A more complex method to evaluate the price of the option would be to use its expected profits. 
	At a $\$200$ rate, activating the \gls{ASIC} will result in a loss of $\$50$, as $\$250$ is paid and only $\$200$ is received; thus, rational agents will not activate the \gls{ASIC}, and will lose nothing. 
	On the other hand, if the exchange rate increases to $\$800$, it is possible to earn $\$800 - \$250 = \$550$ by turning the hardware on. 
	In total, the expected return is $\frac{1}{2} \cdot \$0 + \frac{1}{2} \cdot \$550 = \$275$. 
	
	It is tempting to say that this is the correct price for the option, but it does not take exchange-rate volatility into account. We later show that the correct price is in fact $\$183\frac{1}{3}$. 
	
	To show why both $\$150$ and $\$275$ are \emph{incorrect}, note that these prices create arbitrage opportunities. We proceed by constructing a trading strategy that capitalizes on the arbitrage created by the latter price, and note that a similar strategy can be used for the former. Assume there is at least one rational buyer for the opportunity, willing to pay $\$275$. If so, that buyer will surely prefer purchasing it for the lower price of $\$274$!
	
	We can sell the opportunity for the lower price \emph{without} actually owning it, all the while promising the buyer that no matter the world state the same exact profits will be earned. Essentially, we are performing a short on the opportunity.
	
	To fulfill the promise we do the following: immediately upon selling the opportunity we borrow $\$183\frac{1}{3}$ from the bank, giving us a total of $\$183\frac{1}{3} + \$274 = \$457\frac{1}{3}$. We buy $\bitcoin \frac{11}{12}$, which under the current exchange-rate are worth $\frac{11}{12} \cdot \$400 = \$366\frac{2}{3}$. After this, we remain with a profit of $\$457\frac{1}{3} - \$366\frac{2}{3} = \$90\frac{2}{3}$, which we pocket as a profit.
	This is summarized in \cref{table:SimpleExampleDayOne}.
	
	If bitcoin's value goes up, our rational buyer will want to turn on the (imaginary) \gls{ASIC} and receive the promised $\bitcoin 1$ reward in exchange for the $\$250$ activation fee, which is paid to us. We use the fee to pay back the loan, leaving us with $\$250 - \$183\frac{1}{3} = \$66\frac{2}{3}$, exactly enough to buy $\bitcoin \frac{1}{12}$, that together with our existing $\bitcoin \frac{11}{12}$ can be given to the buyer as the mining reward, thus covering our short. Note we have also paid back all debt, while our pocketed $\$90\frac{2}{3}$ profit was untouched.
	The balances throughout the day are shown in \cref{table:SimpleExampleDayTwoUpState}.
	
	On the other hand, if the value goes down, the rational buyer will not want to pay the activation fee as it is more expensive than the $\bitcoin 1$ ($= \$200$) profit; even if the buyer is interested in receiving a single bitcoin, buying it on the free market is cheaper than activating the \gls{ASIC}. So, we have covered our short without having to pay the mining reward. We still need to repay our $\$183\frac{1}{3}$ debt, and luckily our coins are worth exactly $\frac{11}{12} \cdot \$200 = \$183\frac{1}{3}$. Again, we keep our pocketed profit. 
	\cref{table:SimpleExampleDayTwoDownState} presents all changes in our holdings.
	
	Although we started with no money, we made a riskless profit of $\$90\frac{2}{3}$ due to the \emph{incorrect} pricing of the \gls{ASIC}. In \cref{sec:Results}, we show how to correctly price it, and prove that when using our method no arbitrage opportunities arise.
	
    \begin{table}
        \centering
        \caption{Balance of all assets on the first day of \cref{exm:SimpleExample}. In step $\#1$, after selling the opportunity we have a $-1$ quantity of it, essentially performing a short on it.}
        \label{table:SimpleExampleDayOne}
        \scalebox{1} {
        \begin{tabular}{c l c c c c}
            \toprule
            
            \# & 
            Step & 
            Cash & 
            Debt & 
            Coins & 
            Opportunities \\
            
            \midrule
            
            0 & 
            Start of day. & 
            $\$0$ & 
            $\$0$ & 
            $0$ & 
            $0$ \\
            
            1 & 
            Sell opportunity. & 
            $\$274$ & 
            $\$0$ & 
            $0$ & 
            $-1$ \\
            
            2 & 
            Borrow $\$183\frac{1}{3}$. &
            $\$457\frac{1}{3}$ &
            $\$183\frac{1}{3}$ & 
            $0$ & 
            $-1$ \\
            
            3 & 
            Buy $\frac{11}{12}$ coins. &
            $\$90\frac{2}{3}$ &
            $\$183\frac{1}{3}$ &
            $\frac{11}{12}$ & 
            $-1$ \\
            
            \bottomrule
        \end{tabular}
        }
    \end{table}
    \begin{table}
        \centering
        \caption{Balance of all assets on the second day of \cref{exm:SimpleExample}, if the exchange-rate has doubled. Regarding step $\#4$: giving the buyer $1$ coin covers the short on the opportunity.}
        \label{table:SimpleExampleDayTwoUpState}
        \scalebox{1} {
        \begin{tabular}{c l c c c c}
            \toprule
            
            \# & 
            Step & 
            Cash & 
            Debt & 
            Coins & 
            Opportunities \\
            
            \midrule
            
            0 & 
            Start of day. &
            $\$90\frac{2}{3}$ &
            $\$183\frac{1}{3}$ &
            $\frac{11}{12}$ & 
            $-1$ \\
            
            1 & 
            Get activation fee. &
            $\$340\frac{2}{3}$ &
            $\$183\frac{1}{3}$ &
            $\frac{11}{12}$ & 
            $-1$ \\
            
            2 & 
            Pay loan back. &
            $\$157\frac{1}{3}$ &
            $\$0$ & 
            $\frac{11}{12}$ & 
            $-1$ \\
            
            3 & 
            Buy $\frac{1}{12}$ coins. &
            $\$90\frac{2}{3}$ &
            $\$0$ & 
            $1$ & 
            $-1$ \\
            
            4 & 
            Pay buyer $1$ coin. &
            $\$90\frac{2}{3}$ &
            $\$0$ & 
            $0$ & 
            $0$ \\
            
            \bottomrule
        \end{tabular}
        }
    \end{table}
    \begin{table}
        \centering
        \caption{Asset balance on the second day, if the exchange-rate has halved. Rational buyers will not activate the \gls{ASIC} for a loss, thus there is a $0$ amount of the opportunity at step $\#0$.
        }
        \label{table:SimpleExampleDayTwoDownState}
        \scalebox{1} {
        \begin{tabular}{c l c c c c}
            \toprule
            
            \# & 
            Step & 
            Cash & 
            Debt & 
            Coins & 
            Opportunities \\
            
            \midrule
            
            0 & 
            Start of day. &
            $\$90\frac{2}{3}$ &
            $\$183\frac{1}{3}$ &
            $\frac{11}{12}$ & 
            $0$ \\
            
            1 & 
            Sell all coins. &
            $\$274$ &
            $\$183\frac{1}{3}$ & 
            $0$ & 
            $0$ \\
            
            2 & 
            Pay loan back. &
            $\$90\frac{2}{3}$ &
            $\$0$ & 
            $0$ & 
            $0$ \\
            
            \bottomrule
        \end{tabular}
        }
    \end{table}
\end{example}

\paragraph*{Organization}
This paper is structured as follows:
we present additional background on cryptocurrencies and option theory in \cref{sec:Background}. We go on to define a mining model in \cref{sec:Model}, and present our methods for correctly pricing \glspl{ASIC} in \cref{sec:Results}, deferring most proofs to \cref{sec:Proofs}.
We then employ our methods to perform an empirical evaluation using real-world data in \cref{sec:EmpiricalEvaluation}.  We go over related work in \cref{sec:RelatedWork} and conclude with a discussion on the implication of our results and future work in \cref{sec:conclusion}.

\section{Background}
\label{sec:Background}
We now go over preliminary details necessary for our work.
We begin by describing in \cref{sec:cryptocurrencies} the mechanisms which underlie \gls{PoW} cryptocurrencies, and by reviewing the economical considerations made by real-world miners in \cref{sec:MiningEconomicalDecision}. We finish by giving a brief overview of option theory in \cref{sec:optionTheory}.

\subsection{Cryptocurrencies}
\label{sec:cryptocurrencies}
Bitcoin and other similar tokens let users exchange funds by creating \emph{transactions} \cite{gafni2022greedy} that are collected in \emph{blocks} in a decentralized manner by pseudonymous users called \emph{miners}, who are allowed to freely join or leave the system. The creation of blocks is called \emph{mining}.
To enforce some chronological order on transactions, each block must point to a preceding one, with the resulting data-structure often referred to as a \emph{blockchain}. Thus, a blockchain is in essence a decentralized ledger of transactions, where blocks should ideally be mined one after the other.

\paragraph*{Proof-of-Work}
Bitcoin relies on a mechanism called \gls{PoW} to ensure miners invest some expected amount of effort to create blocks, thus preventing miners from maliciously retroactively changing the ledger to their benefit \cite{yaish2023uncle}.
This is enforced by requiring blocks to have a cryptographic hash \cite{katz2020introduction} which is lower than some \emph{target} value (when this hash is interpreted as a number).
The hash function used in Bitcoin is \emph{SHA256} \cite{dang2015secure}. Currently, the best known method for finding a low SHA256 hash is to try many different inputs by brute force \cite{katz2020introduction}. Thus, the performance of mining hardware is measured by its \emph{hash-rate}, the amount of hash calculations it can perform per unit of time.
The mining target value is set by the mechanism to keep block creation rate roughly constant even when computational power is added to the network \cite{yaish2022blockchain}.
The specific mechanism overseeing this is called the \emph{\gls{DAA}}.
Thus, the probability that a single miner will create a block decreases if more hash-rate is competing against it.

In our work, we focus on Bitcoin.
Historical data shows that Bitcoin’s hash-rate was consistently more than 100 times higher than the combined power of other popular cryptocurrencies \cite{bitinfocharts2022bitcoin}. This allows us to avoid considerations such as ``coin-hopping'' (wherein miners switch between mining different cryptocurrencies), similarly to other papers \cite{eyal2014majority,yaish2022blockchain,sompolinsky2016bitcoins,garay2015bitcoin,tsabary2018gap}. Indeed, previous research shows that such behavior is rare in practice \cite{long2022measuring}.

\paragraph*{ Mining Incentives}
To encourage mining even in the face of the ever-mounting computational effort required, Bitcoin and similar cryptocurrencies reward the creator of each block with a \emph{block reward}.
As the size of blocks is limited, users can incentivize miners to prefer their transaction over others by paying a \emph{transaction fee} to the first miner that includes it in a block.
Transaction fees have roughly amounted to $1.5\%$ of Bitcoin mining profits over the past year \cite{btc.com2022pool}.

Single miners do not expect to find a block often, and so the majority of bitcoin mining is done in mining \emph{pools} \cite{wang2020measurement,silva2020impact,tovanich2021visual}, where miners mine cooperatively and split rewards amongst themselves according to their relative contribution. Thus, small and constant returns can be expected by miners who take part in pools.

\subsection{Real-World Considerations of Miners}
\label{sec:MiningEconomicalDecision}

Cryptocurrency exchange-rates and electricity costs are important considerations for miners \cite{rauchs2017global,fabiano2022today,naughton2022as,volpicelli2022as}. This is affirmed by large-scale miners, who claim to respond to market changes by turning mining rigs on and off ``at a minute's notice'' and ``in real-time'' \cite{sigalos2022as,helman2020how,news2021bitcoin}. Indeed, historical data indicates that miners rapidly turn hardware on and off, going as far as using old and inefficient hardware when the current rates deem it profitable \cite{harper2022hashrate}.
On the other end of the spectrum, large-scale miners are not afraid of shutting hardware down for prolonged periods of time to move it to remote areas with cheap electricity \cite{redman2019chinese,chipolina2020bitcoins,haig2020btc}.

Such behavior is facilitated by hardware and software vendors, who create products that are designed to rapidly switch between multiple low-power modes according to market conditions \cite{blockstream2021instant,technologies2019building,harper2022hashrate,company2022difference,company2020s19}.

Even amateur miners use such optimizations by adopting after-market software that adds similar functionality to hardware which doesn't have it by default \cite{taserz2019asic.to,braiins2018braiins,antpool2020antminertool,btc.com2019using,vera2020what,braiins2022autotuning,vnish2022firmware,company2022recommended}.

\subsection{Financial Options}
\label{sec:optionTheory}

A European \emph{call-option} is a contract involving two parties and an underlying asset. By purchasing a call-option, the buyer receives from the seller the right to buy the asset at some agreed-upon price, the \emph{strike price}, at a specific future date, the \emph{expiration date}. As this is a right and not an obligation, the buyer need not exercise it if deemed unprofitable. 
For example, if by the date of expiry the underlying asset's price is lower than the strike price, it is preferable to buy the underlying asset directly and to discard the option.

In 1973, the Black-Scholes method for option valuation was proposed by \cite{black1973pricing}, a seminal work in the field of option theory, and was later expanded upon by Merton \cite{merton1973theory}. Both rely on the \emph{no-arbitrage} principle which argues that options should be priced such that no arbitrage possibility involving the underlying asset exists.
Using option pricing as a foundation, various financial decisions have been cast as options \cite{brandao2005using,trigeorgis2017real,trigeorgis1996real,copeland2001real}, for example the decision of whether to delay or abandon a project. This technique is called \emph{real option valuation}.

Techniques from real option theory are introduced as needed throughout \cref{sec:Results}, with the required modifications for our setting. Further exploration of the topic is beyond the scope of this paper, but can be found in classic texts such as \cite{cochrane2009asset}.

\section{Model}
\label{sec:Model}

We now describe an accurate model which accounts for the considerations made by real-world miners (see \cref{sec:MiningEconomicalDecision}).

\subsection{Mining Model}
We divide time into discrete mining \emph{opportunities} (or \emph{turns}), and assumes a miner can either activate its hardware or leave it off for the whole duration of a single turn $\finalturn$.

If the \gls{ASIC} has a hash-rate of $\hashrate$ hashes-per-second and the total hash-rate active on the network excluding the \gls{ASIC} is $\globalhashrate{\finalturn}$, activation of the \gls{ASIC} allows the miner to receive a fraction $\hashratio{\finalturn}$ of the block-reward, which is $\blockreward{\finalturn}$ coins. This is a highly accurate approximation of the rewards earned by mining, as explained previously in \cref{sec:Background}.

Denote the \gls{ASIC}'s efficiency, measured in the \glspl{kWh} required for the computation of a mining opportunity, as $\efficiency$, and the cost of electricity as $\electricityprice{\finalturn}$ dollars per \gls{kWh}.

To model hardware failures, we assume the \gls{ASIC} ``decays'' gradually according to a mortality distribution: let $\mortality \left( \finalturn \right)$ be the fraction of the \gls{ASIC} that ``remains'' after $\finalturn$ time units. For example, $\mortality$ can be a complementary \gls{CDF} of some distribution \cite{jeffreys1998theory}; let $F$ be the \gls{CDF}, then the complementary \gls{CDF} is defined to be $1-F$.

\subsection{Financial Model}
In our financial model of the world, for simplicity we call the mined cryptocurrency ``Bitcoin'', and refer to the fiat currency in which mining expenses are paid for as the \glsxtrshort{USD}, but both can be replaced by any other similar cryptocurrency and fiat currency.

We model the change in Bitcoin's exchange rate as a multiplicative random walk.
We denote the Bitcoin-to-\glsxtrshort{USD} exchange rate at turn $\finalturn$ by $\coinprice{\finalturn}$, the probability for its value to rise to $\upcoinprice \coinprice{\finalturn}$ in the next turn by $\upprobability$, and to fall to $\downcoinprice \coinprice{\finalturn}$ in the next turn by $\downprobability$, resulting in a binomial price tree.
A general form of such a tree is depicted in \cref{fig:PriceTree}.

\begin{figure}
	\centering
	\begin{forest}
		for tree={circle, draw, align=center}
		[{$\coinprice{0}$}
		[{$\dots$}, name=leftdots, no edge, tier=dots]
		[{$\upcoinprice \coinprice{0}$}, name=leftbranch, edge label={node[midway, fill=white] {$\upprobability$}} 
		[{$\upcoinprice ^2 \coinprice{0}$}, name=leftchild, tier=bottom
		[, phantom, tier=dots]
		[, phantom, tier=dots]
		]
		[, phantom, tier=bottom]
		]
		[{$\dots$}, name=leftshareddots, no edge, tier=dots]
		[{$\upcoinprice \downcoinprice \coinprice{0}$}, name=sharedchild, no edge, tier=bottom,
		[, phantom, tier=dots]
		[, phantom, tier=dots]
		]
		[{$\dots$}, name=rightshareddots, no edge, tier=dots]
		[{$\downcoinprice \coinprice{0}$}, name=rightbranch, edge label={node[midway, fill=white] {$\downprobability$}} 
		[, phantom, tier=bottom]
		[{$\downcoinprice ^ 2 \coinprice{0}$}, name=rightchild, tier=bottom
		[, phantom, tier=dots]
		[, phantom, tier=dots]
		]
		]
		[{$\dots$}, name=rightdots, no edge, tier=dots]
		\draw (leftbranch)--(sharedchild);
		\draw (rightbranch)--(sharedchild);
		\draw (leftchild)--(leftdots);
		\draw (leftchild)--(leftshareddots);
		\draw (sharedchild)--(leftshareddots);
		\draw (sharedchild)--(rightshareddots);
		\draw (rightchild)--(rightshareddots);
		\draw (rightchild)--(rightdots);
		]
	\end{forest}
	\caption{A coin's exchange rate as a multiplicative random walk, with a start value of $\coinprice{0}$, a $\upprobability$ probability to increase by a factor of $\upcoinprice$, and a $\downprobability$ probability to decrease by $\downcoinprice$.}
	\label{fig:PriceTree}
\end{figure}

While it may seem simplistic to assume that the price at every step can either increase or decrease by a factor, using sufficiently small steps yields a granular price model. Indeed, this distribution is commonly used in finance to model the value of assets such as currencies and stocks \cite{brandao2005using,rubinstein1994implied}. Note that the length of each step of the exchange-rate's random walk does not have to coincide with the length of a mining opportunity. As we focus on evaluating a single mining opportunity, we use arbitrarily small steps to achieve a high granularity.

Denote the economy's annual multiplicative risk-fee rate as $\riskfreerate$. We assume $0 < \downcoinprice < 1 < \riskfreerate < \upcoinprice$, otherwise, riskless arbitrage opportunities emerge, which we assume to not exist. This assumption is crystallized in \cref{definition:noArbitrage}.

\begin{definition}[The no-arbitrage assumption]
    \label{definition:noArbitrage}
	The free market adjusts asset prices such that it is impossible to outpace market gains without exposure to more risk. If such an arbitrage opportunity arises, market forces quickly use it until a pricing equilibrium is found, thus closing the opportunity.
\end{definition}

We mainly deal with the following types of assets:
\begin{enumerate}
	\item[i.] The underlying cryptocurrency.
	\item[ii.] A mining opportunity, denoting its value as $\optionvalue{\cdot}$.
	\item[iii.] A risk-free asset. An asset with a future return which is independent of the state of the world that is reached. Its multiplicative return is the \emph{risk-free rate}. An example of such an asset is a government-issued bond, the value of which is denoted by $\bonds{}$. 
\end{enumerate}

We also create portfolios holding combinations of the above assets, and denote their values by $\portfoliovalue{\cdot}$.
We assume that assets are traded with sufficient liquidity, a clearly defined price and that it is possible to hold a ``short'' position on each one (owing the asset to another party, equivalent to holding a negative amount of it).

\section{Theoretical Results}
\label{sec:Results}

In this section, we derive the main results that allow us to evaluate the price of a mining machine, with all proofs given in \cref{sec:Proofs}.

\subsection{Pricing an ASIC}
An \gls{ASIC} gives its owner an option to activate it for each of the mining opportunities available during its lifetime, so an \gls{ASIC}'s value is exactly the sum of the values of all these opportunities, and by pricing a single opportunity we can price an \gls{ASIC}.

\begin{definition}[The value of the $\finalturn$-th mining opportunity, at turn $\firstturn$.]
    \label{definition:opportunityValue}
    Let $\firstturn \le \finalturn$. Given that the coin's exchange rate at $\firstturn$ is $\coinprice{\firstturn}$, we shall denote the value of the $\finalturn$-th opportunity at time $\firstturn$ as $\optionvalue{\finalturn, \firstturn, \coinprice{\firstturn}}$.
\end{definition}
Some parameters (such as $\hashrate$) are left out of the notation for brevity.

Recall \cref{exm:SimpleExample}, which has demonstrated that it is hard to evaluate a future mining opportunity, e.g. calculate $\optionvalue{\finalturn, \firstturn, \coinprice{\firstturn}}$ when $\firstturn < \finalturn$, as the future exchange-rate is unknown. Specifically, that example examined a very basic case: $\optionvalue{1, 0, \$400}$. Thus, we take a step back and attempt to evaluate something easier, starting with each option's ``immediate'' value, which we soon define, and use a series of theorems and claims to evaluate a future option's value relative to arbitrary points in time, thereby giving the tools to calculate $\optionvalue{\finalturn, \firstturn, \coinprice{\firstturn}}$.

\paragraph*{Total \gls{ASIC} Value}
Assuming we have successfully evaluated \gls{ASIC} activation for a single turn, we can evaluate an ``entire'' \gls{ASIC} received at time $\asicturn$, relative to time $\finalturn \le \asicturn$:
\begin{equation}
    \label{eq:AsicValue}
	\asicvalue{\asicturn, \finalturn, \coinprice{\finalturn}} \define \sum_{\turnReg=\asicturn}^\infty \mortality \left( \turnReg - \asicturn \right) \cdot \optionvalue{\turnReg, \finalturn, \coinprice{\finalturn}}
\end{equation}

\paragraph*{Reception Delay}
A method for evaluating \gls{ASIC} prices could allow us to estimate the potential decrease in price associated with receiving hardware farther in the future.
Often, \gls{ASIC} manufacturers are backlogged and either deliver orders in the far future, or charge a premium for early deliveries. Assuming \glspl{ASIC} do not decay while in transit, the loss of receiving an \gls{ASIC} at time $\asicturn'$ instead of $\asicturn$ is: 
\begin{equation}
    \label{eq:ReceptionDelay}
	\asicvalue{\asicturn', \finalturn, \coinprice{\finalturn}} - \asicvalue{\asicturn, \finalturn, \coinprice{\finalturn}}
\end{equation}

\subsection{Pricing the Current Mining Opportunity}

We begin by evaluating the $\finalturn$-th opportunity relative to turn $\finalturn$. Following \cref{definition:opportunityValue}, this is notated by $\optionvalue{\finalturn, \finalturn, \coinprice{\finalturn}}$.
At turn $\finalturn$, we know everything required to calculate the value of the $\finalturn$-th mining opportunity, as the biggest cause of uncertainty, the cryptocurrency's exchange rate $\coinprice{\finalturn}$, is given. Thus, we call $\optionvalue{\finalturn, \finalturn, \coinprice{\finalturn}}$ the \emph{immediate} value of the $\finalturn$-th mining opportunity.

\paragraph*{Immediate Value of a Single Opportunity}
At the $\finalturn$-th  mining opportunity, an \gls{ASIC}'s owner has the option of paying the electricity cost of activating the \gls{ASIC} for the duration of the opportunity, which under our model is $\hashrate \cdot \efficiency \cdot \electricityprice{\finalturn}$, and in return receive the partial reward of $\frac{\hashrate}{\globalhashrate{\finalturn} + \hashrate} \cdot \blockreward{\finalturn} \cdot \coinprice{\finalturn}$.
This opportunity can never be worth strictly less than zero, as a miner is not obliged to turn on its \gls{ASIC}, and indeed a rational miner will not do so if it incurs a loss.

In total, the value at time $\finalturn$ of the $\finalturn$-th mining opportunity is:
\begin{equation}
    \label{eq:oppValImmediate}
	\optionvalue{\finalturn, \finalturn, \coinprice{\finalturn}} \define \max \left( \frac{\hashrate}{\globalhashrate{\finalturn} + \hashrate} \blockreward{\finalturn} \coinprice{\finalturn} - \hashrate \efficiency \electricityprice{\finalturn}, 0 \right)
\end{equation}

\paragraph*{Shutdown Price}
Immediately arising from \cref{eq:oppValImmediate} is that if the cost of turning on the \gls{ASIC} exceeds the profits, meaning that $\frac{\hashrate}{\globalhashrate{\finalturn} + \hashrate} \blockreward{\finalturn} \coinprice{\finalturn} \le \hashrate \efficiency \electricityprice{\finalturn}$, then no miner will turn it on, as paying the activation cost to buy the mined cryptocurrency on the free market is a better deal than actually using the hardware.
This corresponds with the behavior of actual miners, as described in \cref{sec:MiningEconomicalDecision}.

\subsection{Pricing the Next Mining Opportunity}
\label{sec:PricingSingleMiningOpportunityRelativeToOneTurn}
We now tackle the problem presented in the previous section more generally -- pricing the $\finalturn$-th mining opportunity in relation to turn $\finalturn-1$. We do so by modifying techniques from option-pricing theory (as in \cite{black1973pricing,cox1979option}). Specifically, to price this mining opportunity, we construct a portfolio of mining opportunities and coins at turn $\finalturn-1$.

The portfolio is crafted to yield identical valuations at turn $\finalturn$ regardless of the change in the exchange-rate (see \cref{clm:ShortAmount}).
Thus, it is termed a \emph{risk-free} portfolio.
Its exact value at $\finalturn-1$ can be known by discounting and accounting for the risk-free rate (see \cref{thm:RiskFreePortfolioReturn}).

We consider a portfolio that consists of the $\finalturn$-th mining opportunity and a short on (a yet to be chosen amount of) $\shortcoins{\finalturn-1}$ coins, thus its value at turn $\finalturn-1$ is:
\begin{equation}
    \label{eq:PortfolioValPreviousTurn}
	\portfoliovalue{\finalturn-1} = \optionvalue{\finalturn, \finalturn-1, \coinprice{\finalturn-1}} - \shortcoins{\finalturn-1} \coinprice{\finalturn-1}
\end{equation}
And, its value at turn $\finalturn$ is:
\begin{equation}
    \label{eq:PortfolioValCurrentTurn}
	\portfoliovalue{\finalturn} = \optionvalue{\finalturn, \finalturn, \coinprice{\finalturn}} - \shortcoins{\finalturn-1} \coinprice{\finalturn}
\end{equation}

\begin{restatable}{claim}{claimShortAmount} \label{clm:ShortAmount}
	A portfolio holding the $\finalturn$'th mining opportunity and a short on $\shortcoins{\finalturn-1}$ coins, where:
	    $\shortcoins{\finalturn-1} = \frac{\optionvalue{\finalturn, \finalturn, \upcoinprice \coinprice{\finalturn-1}} - \optionvalue{\finalturn, \finalturn, \downcoinprice \coinprice{\finalturn-1}}}{\coinprice{\finalturn-1} \left( \upcoinprice - \downcoinprice \right)}$.
	is a risk free-portfolio for the turn between $\finalturn-1, \finalturn$. The portfolio's value in all possible states at $\finalturn$ is: 
	$\portfoliovalue{\finalturn} 
	= \optionvalue{\finalturn, \finalturn, \upcoinprice \coinprice{\finalturn-1}} - \shortcoins{\finalturn-1} \upcoinprice \coinprice{\finalturn-1}$.
\end{restatable}	
\cref{sec:Proofs} contains a formal proof.
The main idea is that there is one degree of freedom (choosing the short amount, $\shortcoins{\finalturn-1}$) which must satisfy an equation equating the value of the portfolio in both possible world states.

We now evaluate the return of the portfolio, and use it to price the mining opportunity. 

\begin{restatable}{theorem}{theoremRiskFreePortfolioReturn} \label{thm:RiskFreePortfolioReturn}
	If no arbitrage opportunities exist, the multiplicative return of holding the portfolio constructed in \cref{clm:ShortAmount} between turns $\finalturn-1$ and $\finalturn$ is equal to the risk-free rate.
\end{restatable}

The proof (given in \cref{sec:Proofs}) shows that every other return contradicts the no-arbitrage assumption.
As in \cref{exm:SimpleExample}, we can make a risk-free profit whenever such arbitrage opportunities arise.

We now reach an expression for the opportunity's price:
\begin{restatable}{corollary}{corollarySingleOpportunityValue} \label{cly:SingleOpportunityValue}
	The value of the $\finalturn$-th opportunity at $\finalturn-1$ is:
	\begin{align*}
    	\optionvalue{\finalturn, \finalturn-1, \coinprice{\finalturn-1}}
        = 
    	\frac{\optionvalue{\finalturn, \finalturn, \upcoinprice \coinprice{\finalturn-1}} - \optionvalue{\finalturn, \finalturn, \downcoinprice \coinprice{\finalturn-1}}}{\upcoinprice - \downcoinprice} \left( 1 - \frac{\upcoinprice}{\riskfreerate} \right)
        + 
        \frac{\optionvalue{\finalturn, \finalturn, \upcoinprice \coinprice{\finalturn-1}}}{\riskfreerate}
	\end{align*}
	In the above, all factors can be calculated at time $\finalturn-1$. 
\end{restatable}
We provide a proof in \cref{sec:Proofs}.
It consists of using the return of the portfolio together with its values at turns $\finalturn-1$ and $\finalturn$ to extract the value of the opportunity at $\finalturn-1$.

In \cref{exm:SingleOpportunityValue}, we revisit \cref{exm:SimpleExample} and apply \cref{cly:SingleOpportunityValue} to it.
\begin{example}
    \label{exm:SingleOpportunityValue}
	Surprisingly, the price of the opportunity shown in \cref{exm:SimpleExample} is lower than the na\"ive estimate.
	The opportunity's immediate value if the exchange-rate has gone up is:
	\begin{equation*}
    	\optionvalue{1, 1, 800} = \max{\left( 1 \cdot 800 - 250, 0 \right)} = \$550
	\end{equation*}
	And, for the down state it is:
	\begin{equation*}
	    \optionvalue{1, 1, 200} = \max{\left( 1 \cdot 200 - 250, 0 \right)} = \$0
	\end{equation*}
	By plugging the above into \cref{cly:SingleOpportunityValue} we obtain the correct value of the opportunity at turn $0$:
	\begin{equation*}
	    \optionvalue{1, 0, 400} = \frac{550}{1} + \frac{550 - 0}{2 - \frac{1}{2}} \left( 1 - \frac{2}{1} \right) = \$183\frac{1}{3}
	\end{equation*}
	According to \cref{thm:RiskFreePortfolioReturn}, any other price creates arbitrage.
\end{example}

\subsection{Pricing Relative to an Arbitrary Time} \label{sec:PricingRelativeToArbitraryTime}
\cref{alg:MiningOpportunityValue} extends the previous method to evaluate the $\finalturn$-th opportunity relative to any previous point in time $\firstturn$.
The idea behind the algorithm is to apply the same methods of \cref{sec:PricingSingleMiningOpportunityRelativeToOneTurn} on every possible world-state, starting from turn $\finalturn$ and going back, one step at a time, until reaching $\firstturn$. We now proceed to explain the method in depth.

\begin{algorithm}
	\caption{MiningOpportunityValue}
	\label{alg:MiningOpportunityValue}
	\SetKwInOut{Input}{Input}
	\SetKwInOut{Output}{Output}
	\Input{
		$\finalturn$ - the mining opportunity to evaluate.
		\\
		$\firstturn$ - the turn to evaluate relative to.
		\\
		$\coinprice{\firstturn}$ - coin's exchange-rate at turn $\firstturn$.
	}
	\Output{
		value of $\finalturn$-th opportunity at turn $\firstturn$.
	}
	\For{$\coinprice{\finalturn} \in \{ \upcoinprice^{\finalturn-\firstturn} \cdot \coinprice{\firstturn}, \upcoinprice^{\finalturn-\firstturn-1} \cdot \downcoinprice \cdot \coinprice{\firstturn}, \dots, \downcoinprice^{\finalturn-\firstturn} \cdot \coinprice{\firstturn} \}$}
	{
		$ \optionvalue{\finalturn, \finalturn, \coinprice{\finalturn}} \leftarrow \hashrate \cdot \max \left( \frac{\blockreward{\finalturn} \cdot \coinprice{\finalturn}}{\globalhashrate{\finalturn} + \hashrate} - \efficiency \cdot \electricityprice{\finalturn}, 0 \right) $ \\
	}
	\For{$\turnReg \in \finalturn-1, \dots, \firstturn$}
	{
		\For{$\coinprice{\turnReg} \in \{\upcoinprice^{\turnReg} \coinprice{\firstturn}, \upcoinprice^{\turnReg-1} \downcoinprice \coinprice{\firstturn}, \dots, \upcoinprice \downcoinprice^{\turnReg-1} \coinprice{\firstturn}, \downcoinprice^{\turnReg}  \coinprice{\firstturn} \}$}
		{
			$\shortcoins{\turnReg} \leftarrow \frac{\optionvalue{\finalturn, \turnReg+1, \upcoinprice \cdot \coinprice{\turnReg}} - \optionvalue{\finalturn, \turnReg+1, \downcoinprice \cdot \coinprice{\turnReg}}}{\coinprice{\turnReg} \cdot \left( \upcoinprice - \downcoinprice \right)}$ \\
			$\portfoliovalue{\turnReg+1} \leftarrow \optionvalue{\finalturn, \turnReg+1, \upcoinprice \cdot \coinprice{\turnReg}} - \shortcoins{\turnReg} \cdot \upcoinprice \cdot \coinprice{\turnReg}$ \\
			$\optionvalue{\finalturn, \turnReg, \coinprice{\turnReg}} \leftarrow \shortcoins{\turnReg} \cdot \coinprice{\turnReg} + \frac{\portfoliovalue{\turnReg+1}} {\riskfreerate}$ \\
		}
	}
	\Return $\optionvalue{\finalturn, \firstturn, \coinprice{\firstturn}}$ \\
\end{algorithm}

The random-walk describing the coin's exchange rate for the period between turns $\firstturn$ and $\finalturn$ forms a tree with root $\coinprice{\firstturn}$ and leaves $\upcoinprice^\turnReg \downcoinprice ^{\finalturn - \firstturn - \turnReg} \coinprice{\firstturn}$, for every $\turnReg \in \left[ 0, \finalturn - \firstturn \right]$. 
The leaves represent the trivial cases for evaluation, each one corresponds to a possible world state at turn $\finalturn$. As the opportunity expires at that turn, its value can be calculated directly from the definition given in \cref{eq:oppValImmediate}. 

Proceeding inductively, let $\turnReg \in \left [ \firstturn, \finalturn - 1 \right ]$. We shall evaluate the opportunity at one of the vertices of the $\left( \turnReg - \firstturn \right)$-th level, assume it is $\coinprice{\turnReg}$. It points to two vertices from level $\turnReg - \firstturn + 1$, specifically $\upcoinprice \coinprice{\turnReg}, \downcoinprice \coinprice{\turnReg}$. \cref{sec:PricingSingleMiningOpportunityRelativeToOneTurn} suggests that if the opportunity values for these two vertices are already calculated, the opportunity's value at $\coinprice{\turnReg}$'s world-state can be obtained. \cref{clm:RiskFreePortfolioArbitraryTurn} covers this case.

\begin{restatable}{claim}{claimRiskFreePortfolioArbitraryTurn} \label{clm:RiskFreePortfolioArbitraryTurn}
	Let $\turnReg < \finalturn$. Given that the opportunity's valuations at $\turnReg+1$ are known, it is possible to evaluate $\optionvalue{\finalturn, \turnReg, \coinprice{\turnReg}}$, which is equal to:
	\begin{align*}
    	\optionvalue{\finalturn, \turnReg, \coinprice{\turnReg}}
    	&=
    	\frac{\optionvalue{\finalturn, \turnReg+1, \upcoinprice \coinprice{\turnReg}} - \optionvalue{\finalturn, \turnReg+1, \downcoinprice \coinprice{\turnReg}}}{\upcoinprice - \downcoinprice} \left( 1 - \frac{\upcoinprice}{\riskfreerate} \right)
    	\\ &+
    	\frac{\optionvalue{\finalturn, \turnReg+1, \upcoinprice \coinprice{\turnReg}}}{\riskfreerate}
	\end{align*}
\end{restatable}
The proof is given in \cref{sec:Proofs}.
It uses the valuations at $\turnReg+1$ to create a risk-free portfolio at turn $\turnReg$ that holds the $\finalturn$-th opportunity. The return of the portfolio at $\turnReg+1$ can then be used to retrieve the value of the opportunity, similarly to \cref{cly:SingleOpportunityValue}.

By applying \cref{clm:RiskFreePortfolioArbitraryTurn} on every vertex of the current level and continuing in a dynamic manner to previous levels, it is possible to reach our goal and finally derive the value at the root of the tree, which corresponds to turn $\firstturn$. 

\paragraph*{A formula for a mining opportunity's value}
Careful mathematical reasoning can be applied to \cref{alg:MiningOpportunityValue} to derive a formula for the value of the $\finalturn$-th opportunity:

\begin{restatable}{theorem}{theoremOppValueEquation} \label{thm:oppValueEquation}
	Let $\downcoefficient = \frac{1 - \frac{\upcoinprice} {\riskfreerate}}{\upcoinprice - \downcoinprice}$, $ \upcoefficient = \downcoefficient + \frac{1}{\riskfreerate} $, 
	$
    	\turnReg_0 = \ceil*{
    		\frac
    		{\log \left( 
    			\frac{\left( \globalhashrate{\finalturn} +\hashrate \right) \efficiency \electricityprice{\finalturn}}{\blockreward{\finalturn}  \downcoinprice^{\finalturn - \firstturn} \coinprice{\firstturn}}
    			\right)}
    		{\log \left( \frac{\upcoinprice}{\downcoinprice} \right)}
    	}
	$.
    The value of the $\finalturn$-th mining opportunity at turn $\firstturn < \finalturn$ is: 
 	\begin{equation*}
    	\optionvalue{\finalturn, \firstturn, \coinprice{\firstturn}}  
    	= 
    	\sum_{\turnReg=\turnReg_0}^{\finalturn - \firstturn}  
    	\frac{\binom{\finalturn - \firstturn}{\turnReg} \upcoefficient ^ \turnReg}
    	{\left( - \downcoefficient \right) ^{\firstturn+\turnReg-\finalturn}}
    	\optionvalue{\finalturn, \finalturn, \upcoinprice^{\turnReg}  \downcoinprice^{\finalturn-\firstturn-\turnReg} \coinprice{\firstturn}} 
 	\end{equation*}
\end{restatable}

The proof is given in \cref{sec:Proofs}.
By recursively applying \cref{clm:RiskFreePortfolioArbitraryTurn} on $\optionvalue{\finalturn, \firstturn, \coinprice{\firstturn}}$, a sum that only includes values of immediate opportunities is reached; this sum is shortened by ignoring opportunities with zero value. By \cref{thm:RiskFreePortfolioReturn}, the value which is obtained is the only one which does not violate the no-arbitrage principle.

\cref{exm:twoTurnExample} shows how to use \cref{thm:oppValueEquation} in a complex setting.
\begin{example}
    \label{exm:twoTurnExample}
	Assume that bitcoin's exchange-rate at turn $0$ is $\$200$, and can either double or halve with equal probability. Extending the walk to two turns produces the tree in \cref{fig:ExampleTwoTurnsRandomWalk}.
	
	\begin{figure*}
		\centering
        \begin{subfigure}[b]{0.49\textwidth}
            \centering
    		\scalebox{0.80}{
            \fontsize{10pt}{10pt}
    		\begin{forest}
    			for tree={circle, draw, minimum size=1.4cm, align=center}
    			[{$\$200$}
    			[{$\$400$}, name=leftbranch, edge label={node[midway, fill=white] {$0.5$}} 
    			[{$\$800$}, name=leftchild, tier=bottom
    			]
    			[, phantom, tier=bottom]
    			]
    			[{$\$200$}, name=sharedchild, no edge, tier=bottom,
    			]
    			[{$\$100$}, name=rightbranch, edge label={node[midway, fill=white] {$0.5$}}
    			[, phantom, tier=bottom]
    			[{$\$50$}, name=rightchild, tier=bottom
    			]
    			]
    			\draw (leftbranch)--(sharedchild);
    			\draw (rightbranch)--(sharedchild);
    			]
    		\end{forest}
     		}
    		\caption{\cref{exm:twoTurnExample}'s equiprobable two turn random walk, with a starting exchange-rate of $\$200$ per \glsxtrshort{BTC}.}
    		\label{fig:ExampleTwoTurnsRandomWalk}
        \end{subfigure}
        \hfill
        \begin{subfigure}[b]{0.49\textwidth}
    		\centering
    		\scalebox{0.80}{
            \fontsize{10pt}{10pt}
    		\begin{forest}
    			for tree={circle, draw, minimum size=1.4cm, align=center}
    			[{$\$\frac{550}{9}$}
    			[{$\$\frac{550}{3}$}, name=leftbranch 
    			[{$\$550$}, name=leftchild, tier=bottom
    			]
    			[, phantom, tier=bottom]
    			]
    			[{$\$0$}, name=sharedchild, no edge, tier=bottom,
    			]
    			[{$\$0$}, name=rightbranch
    			[, phantom, tier=bottom]
    			[{$\$0$}, name=rightchild, tier=bottom
    			]
    			]
    			\draw (leftbranch)--(sharedchild);
    			\draw (rightbranch)--(sharedchild);
    			]
    		\end{forest}
    		}
    		\caption{The value of \cref{exm:twoTurnExample}'s mining opportunity at each possible world state, according to \cref{alg:MiningOpportunityValue}.}
    		\label{fig:ExampleTwoTurnsOppValue}
		\end{subfigure}
		\caption{Visual depictions of the possible states for \cref{exm:twoTurnExample}'s exchange-rate random walk, and the corresponding mining opportunity values and imitating portfolios.}
	\end{figure*}
	
	Furthermore, assume the vendor from \cref{exm:SimpleExample} offers you the option of using its ASIC at the second turn for 10 minutes, under the same conditions as before.
	By following \cref{alg:MiningOpportunityValue}, the value of the opportunity at each state can be calculated, as shown in \cref{fig:ExampleTwoTurnsOppValue}. 
	The algorithm proceeds as follows: 
	
	We start from the leaves and evaluate the immediate value of the opportunity at each one. At the leaf where the exchange-rate is $\$800$, the opportunity is worth $\$550$. On the other hand, if the rate is either $\$200$ or $\$50$, the opportunity is worth $\$0$. We have determined the value of the opportunity at all possible states of turn 2.
	
	Now, by using \cref{clm:RiskFreePortfolioArbitraryTurn} on each of the two possible states at turn $1$, we get that the value of the opportunity can be either $\$\frac{550}{3}$ (if the exchange rate is $\$400$) or $\$0$ (if it is $\$100$).
	
	Finally, we take one step back and look at turn $0$. By employing \cref{clm:RiskFreePortfolioArbitraryTurn} again together with our previous results, we find that the opportunity is worth $\$\frac{550}{9}$ at the first turn.
\end{example}

\subsection{Imitating Portfolio}
\label{sec:ImitatingPortfolio}
Buying mining hardware can entail difficulties: storing and maintaining it is costly, and receiving ordered \glspl{ASIC} promptly requires paying a hefty premium when demand is high.

Imitating an \gls{ASIC}'s revenue using purely financial means (e.g., an investment portfolio of tokens) might be better -- it can start to produce revenue immediately without waiting, and avoids the aforementioned expenses. In \cref{thm:ImitatingPortfolio}, we show construct such a portfolio using coins and bonds.

\begin{restatable}{theorem}{theoremImitatingPortfolio}
    \label{thm:ImitatingPortfolio}
	At turn $\turnReg$, it is possible to construct an \emph{imitating portfolio} for the $\finalturn$-th mining opportunity which is comprised of tokens and bonds. If this portfolio is properly adjusted at each turn until reaching time $\finalturn$, it can be sold to produce the same profits as the imitated mining opportunity.
\end{restatable}

The proof relies on a series of claims, which we go over now.
The portfolio we construct imitates the $\finalturn$-th opportunity between turns $\turnReg, \turnReg+1$, for $\turnReg < \finalturn$. Denote by $\longcoins{\turnReg}, \bonds{\turnReg}$ the respective amount of coins and risk-free bonds in the imitating portfolio at time $\turnReg$. Thus, the portfolio's value at time $\turnReg$ is:
\begin{equation}
    \label{eq:ImitatingportfolioValCurrentTurn}
    \imitatingportfoliovalue{\turnReg}
    =
    \bonds{\turnReg} + \longcoins{\turnReg} \cdot \coinprice{\turnReg}
\end{equation}
And, at $\turnReg+1$ it is 
\begin{equation}
    \label{eq:ImitatingPortfolioValNextTurn}
    \imitatingportfoliovalue{\turnReg+1}
    =
    \riskfreerate \cdot \bonds{\turnReg} + \longcoins{\turnReg} \cdot \coinprice{\turnReg+1}
\end{equation}

\begin{restatable}{claim}{claimLongAmount}
    \label{clm:LongAmount}
	If there are no fees for trading bonds and coins, a portfolio can be constructed at turn $\turnReg$ to be worth exactly the same as the $\finalturn$-th mining opportunity in all world-states of turn $\turnReg+1$:
	$
	    \imitatingportfoliovalue{\turnReg+1} = \optionvalue{\finalturn, \turnReg+1, \coinprice{\turnReg+1}}
	$.
    This portfolio is comprised of $\longcoins{\turnReg}$ tokens and $\bonds{\turnReg}$ risk-free bonds, where:
    \begin{align*}
    	\longcoins{\turnReg}
    	&=
    	\frac{\optionvalue{\finalturn, \turnReg+1, \upcoinprice \cdot \coinprice{\turnReg}} - \optionvalue{\finalturn, \turnReg+1, \downcoinprice \cdot \coinprice{\turnReg}}}{\coinprice{\turnReg} \cdot \left( \upcoinprice - \downcoinprice \right)}
	    \\
    	\bonds{\turnReg}
    	&=
    	\frac{\upcoinprice \cdot \optionvalue{\finalturn, \turnReg+1, \downcoinprice \cdot \coinprice{\turnReg}} - \downcoinprice \cdot \optionvalue{\finalturn, \turnReg+1, \upcoinprice \cdot \coinprice{\turnReg}}}{\riskfreerate \cdot \left( \upcoinprice - \downcoinprice \right)}
	\end{align*}
\end{restatable}

The proof is similar to that of \cref{clm:ShortAmount}, see \cref{sec:Proofs} for details.

\begin{restatable}{claim}{claimImitatingPortfolio}
    \label{clm:ImitatingPortfolio}
	At turn $\turnReg$, the portfolio constructed in \cref{clm:LongAmount} is equal in value to the $\finalturn$-th mining opportunity:
	    $\imitatingportfoliovalue{\turnReg} = \optionvalue{\finalturn, \turnReg, \coinprice{\turnReg}}$.
\end{restatable}

The proof is given in \cref{sec:Proofs}
It relies on showing that at turn $\turnReg$ the risk-free portfolio of \cref{clm:RiskFreePortfolioArbitraryTurn} is equal in value to $\bonds{\turnReg}$. Finally, the claim is reached by applying algebraic manipulations to the definitions of the risk-free portfolio and the portfolio of \cref{clm:LongAmount}.

We finish the proof of \cref{thm:ImitatingPortfolio} by combining \cref{clm:LongAmount,clm:ImitatingPortfolio}.
\begin{restatable}{corollary}{corollaryImitating}
    \label{cly:Imitating}
    
	The portfolio of \cref{clm:LongAmount} is an imitating portfolio for the $\finalturn$-th mining opportunity between turns $\turnReg$, $\turnReg+1$, meaning the portfolio is equal in value to the opportunity at both turns.
    Additionally, if there are no fees, selling the imitating portfolio for turns $\turnReg, \turnReg+1$ at turn $\turnReg+1$ generates enough money to buy the imitating portfolio for $\turnReg+1, \turnReg+2$.
    Thus, after the initial investment is made, no influx of funds is required to adjust the portfolio between turns, meaning that the initial purchase of the portfolio costs exactly the same as the opportunity that it imitates.
\end{restatable}

Like in \cref{sec:PricingRelativeToArbitraryTime}, the imitating portfolio can be evaluated at multiple time periods by dynamically moving backwards in time. \cref{alg:ImitateMiningOpportunity} provides an algorithmic construction of such a portfolio.
If the portfolio changes between turns, the necessary adjustments cost additional fees; these are included in the empirical evaluation given in \cref{sec:EmpiricalEvaluation}.

\begin{algorithm}
	\caption{ImitateMiningOpportunity}
	\label{alg:ImitateMiningOpportunity}
	\SetKwInOut{Input}{Input}
	\SetKwInOut{Output}{Output}
	\Input{
		$\finalturn$ - the mining opportunity to imitate.
		\\
		$\firstturn$ - the turn to at which to create the portfolio.
		\\
		$\coinprice{\firstturn}$ - coin's exchange-rate at turn $\firstturn$.
	}
	\Output{
		an imitating portfolio for the $\finalturn$-th opportunity relative to turn $\firstturn$.
	}
	$\longcoins{\finalturn} \leftarrow 0$ \\
	$\bonds{\finalturn} \leftarrow 0$ \\
	\For{$\coinprice{\finalturn} \in \{ \upcoinprice^{\finalturn-\firstturn} \cdot \coinprice{\firstturn}, \upcoinprice^{\finalturn-\firstturn-1} \cdot \downcoinprice \cdot \coinprice{\firstturn}, \dots, \downcoinprice^{\finalturn-\firstturn} \cdot \coinprice{\firstturn} \}$}
	{
		$ \optionvalue{\finalturn, \finalturn, \coinprice{\finalturn}} \leftarrow \hashrate \cdot \max \left( \frac{\blockreward{\finalturn} \cdot \coinprice{\finalturn}}{\globalhashrate{\finalturn} + \hashrate} - \efficiency \cdot \electricityprice{\finalturn}, 0 \right) $ \\
	}
	\For{$\turnReg \in \finalturn-1, \dots, \firstturn$}
	{
		\For{$\coinprice{\turnReg} \in \{\upcoinprice^{\turnReg} \coinprice{\firstturn}, \upcoinprice^{\turnReg-1} \downcoinprice \coinprice{\firstturn}, \dots, \upcoinprice \downcoinprice^{\turnReg-1} \coinprice{\firstturn}, \downcoinprice^{\turnReg} \coinprice{\firstturn} \}$}
		{
		    $\longcoins{\turnReg} \leftarrow \frac{\optionvalue{\finalturn, \turnReg+1, \upcoinprice \cdot \coinprice{\turnReg}} - \optionvalue{\finalturn, \turnReg+1, \downcoinprice \cdot \coinprice{\turnReg}}}{\coinprice{\turnReg} \cdot \left( \upcoinprice - \downcoinprice \right)}$ \\
    	    $\bonds{\turnReg} \leftarrow \frac{\upcoinprice \cdot \optionvalue{\finalturn, \turnReg+1, \downcoinprice \cdot \coinprice{\turnReg}} - \downcoinprice \cdot \optionvalue{\finalturn, \turnReg+1, \upcoinprice \cdot \coinprice{\turnReg}}}{\riskfreerate \cdot \left( \upcoinprice - \downcoinprice \right)}$ \\
    	    $\imitatingportfoliovalue{\turnReg} \leftarrow \bonds{\turnReg} + \longcoins{\turnReg} \coinprice{\turnReg}$ \\
    	    $\optionvalue{\finalturn, \turnReg, \coinprice{\turnReg}} \leftarrow \imitatingportfoliovalue{\turnReg}$ \\
		}
	}
	\Return $\left\{ \left( \longcoins{\turnReg}, \bonds{\turnReg} \right) \vert \, \turnReg \in \firstturn, \dots, \finalturn \right\}$ \\
\end{algorithm}

In \cref{exm:ImitatingPortfolio}, we construct an imitating portfolio using the results of \cref{sec:ImitatingPortfolio}.
\begin{example}
    \label{exm:ImitatingPortfolio}
	
	Recall \cref{exm:twoTurnExample}, we revisit it and construct imitating portfolios for each of the example's states. These portfolios are summarized in \cref{fig:ExampleImitatingPortfolio}.
	Portfolios are comprised of holdings in coins and bonds, thus we represent them as tuples where the left item is the amount of coins, and the right one is the bonds' value in USD. Portfolios are sold on the last turn, so all final portfolios hold no assets.
	The portfolios are constructed like so.
	
	First, evaluate the opportunity's price at all states. Next, apply \cref{clm:LongAmount} on each possible state at turn $1$. The imitating portfolio for the state where the exchange-rate equals $\$400$ is comprised of $\frac{550 - 0}{400 \cdot \left( 2 - 0.5 \right)} = \frac{11}{12}$ coins, and $\frac{2 \cdot 0 - 0.5 \cdot 550}{1 \cdot \left( 2 - 0.5 \right)} = -\$\frac{550}{3}$ worth of bonds.
	On the other hand, if the exchange-rate is $\$100$ then the portfolio has $\frac{0 - 0}{100 \cdot \left( 2 - 0.5 \right)} = 0$ coins and $\frac{2 \cdot 0 - 0.5 \cdot 0}{1 \cdot \left( 2 - 0.5 \right)} = 0$ bonds.
	Finally, the portfolio for the first state has $\frac{\frac{550}{3} - 0}{200 \left( 2 - 0.5 \right)} = \frac{11}{18}$ coins and $\frac{2 \cdot 0 - \frac{1}{2} \cdot \frac{550}{3}}{1 \cdot \left( 2 - 0.5 \right)} = -\frac{550}{9}$ bonds. 
	
	To show that these portfolios are indeed imitating, we analyze their returns on the final turn. If an imitating portfolio is sold on the final turn, by construction its return should equal the one given by the actual mining opportunity. 
	
	If the exchange-rate is $\$800$, the portfolio we constructed is worth $800 \cdot \frac{11}{12} - \frac{550}{3} = \$550$, so selling it produces exactly the same profits as the opportunity at this state. If the exchange-rate is $\$200$, look at the two possible cases: if the previous turn's exchange-rate was $\$400$, our portfolio is comprised of $\frac{11}{12}$ coins and bonds worth $-\$\frac{550}{3}$, thus selling the portfolio gives a profit of $400 \cdot \frac{11}{12} - \frac{550}{3} = \$0$, again equal to the opportunity's. Conversely, if the previous rate was $\$100$, our portfolio holds no assets, so there is nothing to sell, and as before the profit is $\$0$, equal to the opportunity's.
\end{example}

\begin{figure}
    \centering
    \begin{forest}
        for tree={circle, draw, minimum size=1.4cm, inner sep=1pt, align=center}
        [{$\bitcoin\frac{11}{18}, -\$\frac{550}{9}$}
        [{$\bitcoin\frac{11}{12}, -\$\frac{550}{3}$}, name=leftbranch 
        [{$\bitcoin0, \$0$}, name=leftchild, tier=bottom
        ]
        [, phantom, tier=bottom]
        ]
        [{$\bitcoin0, \$0$}, name=sharedchild, no edge, tier=bottom,
        ]
        [{$\bitcoin0, \$0$}, name=rightbranch
        [, phantom, tier=bottom]
        [{$\bitcoin0, \$0$}, name=rightchild, tier=bottom
        ]
        ]
        \draw (leftbranch)--(sharedchild);
        \draw (rightbranch)--(sharedchild);
        ]
    \end{forest}
    \caption{Imitating portfolios for each possible world-state of \cref{exm:ImitatingPortfolio}, per \cref{alg:ImitateMiningOpportunity}.}
    \label{fig:ExampleImitatingPortfolio}
\end{figure}

\section{Empirical Evaluation}
\label{sec:EmpiricalEvaluation}

We now employ our methods on real world data, deriving prices for the \emph{Bitmain Antminer S9}, an \gls{ASIC} which has dominated the market for an extended period of time, and constitutes around $33\%$ of the currently active hash-rate on Bitcoin \cite{harper2022hashrate,volpicelli2022as}.

\subsection{Parameters}
The parameters which are used throughout this section were obtained from real-world data, and were set to the following values:

\paragraph*{ASIC price and specifications}
We compare our prices to historical market prices which were obtained from the manufacturer's Amazon page. We took hardware specification from \cite{value2022asic}, and assumed \glspl{ASIC} last $2$ years on average. In fact, hash-rate considerations imply that their profits vanish even faster.

\paragraph*{Mining and imitation fees}
In our evaluation, we compare between \glspl{ASIC} and their corresponding imitating portfolios. If there were multiple options for relevant parameters, we always chose the ones that make the imitating portfolios' job harder:

\begin{itemize}
    \item \emph{Electricity fees} were set to $\$0.035$ per \gls{kWh}, lower than the average rates paid by industrial users and miners in the US \cite{alternativefinance2022cambridge,administration2022electric}.
    \item \emph{Mining pool fees} were set to $2\%$. Large pools (consistently comprising at least $40\%$ of Bitcoin’s hash-rate over the past year) have asked for $2.5\%$ \cite{gencer2018decentralization,btc.com2022pool,wiki2022comparison}.
    \item \emph{Trading fees} for bonds and \glsxtrshort{BTC}-to-\gls{USD} were set to $1\%$, more than fees offered by large companies. For example, Coinbase asks for $0.6\%$ at most \cite{coinbase2022what}.
\end{itemize}

\paragraph*{Exchange rate, hashrate and interest rate}
The historical \glsxtrshort{BTC}-to-\gls{USD} exchange-rate and global hashrate were taken from \url{blockchain.com}. Annual volatility, defined as the standard deviation of log-returns, and future global hash-rate growth (which we assumed to be exponential in accordance with the literature \cite{bowden2020modeling}) were evaluated using data starting at 2013 and ending at the estimation date. The economy's annual risk-free rate was set to $2\%$.

\subsection{Results}
We now go over the results of our empirical evaluation.

\paragraph*{Official Prices Do Not Account For Risk}
We obtain the correct prices for the Antminer S9 by using \cref{alg:MiningOpportunityValue} with parameters corresponding to various points in time.

\begin{figure}
    \centering
    \includegraphics[trim={0cm 0.0cm 0.0cm 0.0cm},clip,width=0.7\linewidth]{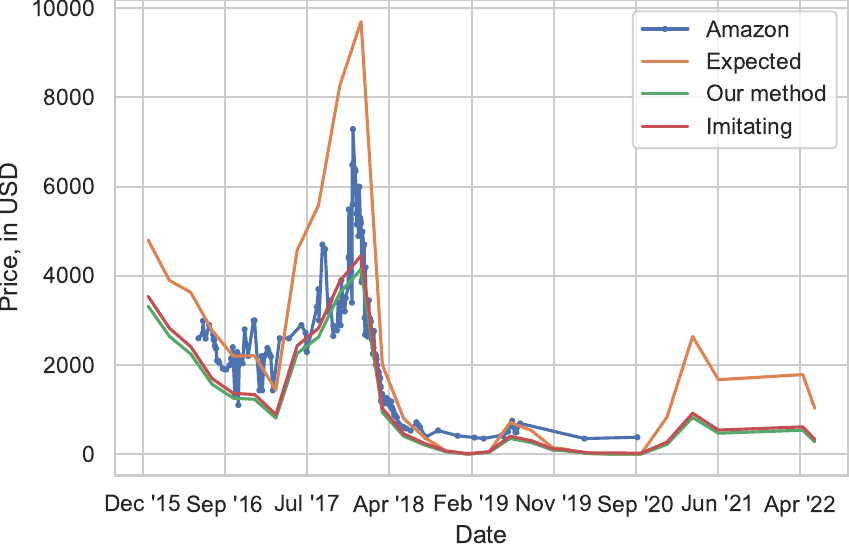}
    \caption{\gls{ASIC} prices according to different valuation methods, as functions of time, including the costs associated with the corresponding imitating portoflios.}
    \label{fig:AsicValueComparison}
\end{figure}

\cref{fig:AsicValueComparison} compares prices given by our method to Bitmain's official Amazon prices, and to a na\"ive evaluation method anecdotally used by miners (labeled ``Expected''), which assumes the future \glsxtrshort{BTC}-\gls{USD} exchange-rate will continue its recent rate of growth. This na\"ive method ignores risk and uses only expected values, as in \cref{exm:SimpleExample}.
The official prices are closer to the na\"ive price, suggesting that \emph{they do not fully account for risk}.

\paragraph*{Imitating Portfolios}
We now utilize \cref{alg:ImitateMiningOpportunity} to produce imitating portfolios for the Antminer S9. These portfolios are benchmarked and compared to the actual hardware using the realized exchange-rates and hash-rates to evaluated the returns made by each.

When evaluated on recent data, our imitating portfolios earned more than the equivalent \glspl{ASIC}, while costing less to buy and maintain, meaning \glspl{ASIC} are overpriced.

\begin{figure}
    \centering
    \includegraphics[trim={0cm 0.0cm 0.0cm 0.0cm},clip,width=0.7\linewidth]{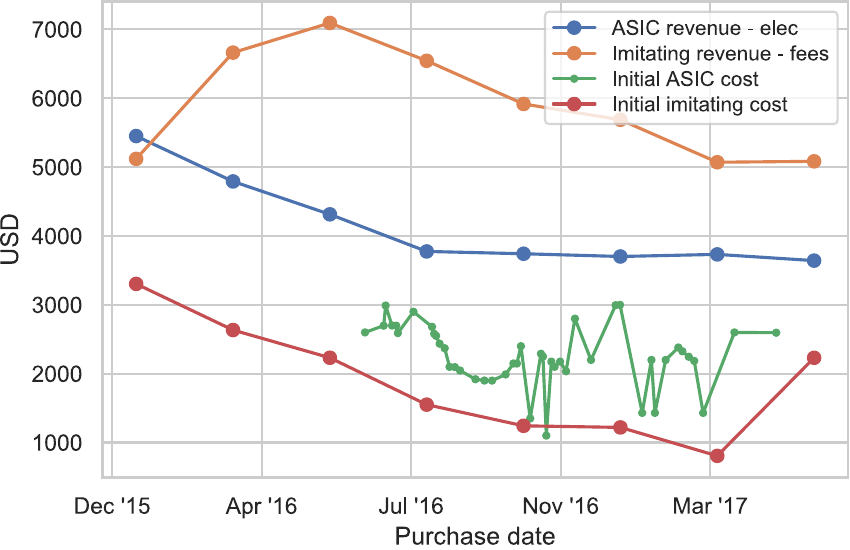}
    \caption{Realized revenue (after expenses) and initial cost for a 2-year operation of an \gls{ASIC} and its corresponding imitating portfolio.
    An \gls{ASIC}'s initial cost is its Amazon price, and its expenses are the electricity it consumes. The portfolio's initial cost is the cost of buying it, and its expenses are the trading fees required for maintaining it over 2 years.
    }
    \label{fig:RevenueComparison}
\end{figure}

\cref{fig:RevenueComparison} aggregates \emph{realized} revenues and initial costs of \glspl{ASIC} and the corresponding imitating portfolios. We assume \glspl{ASIC} are received and activated \emph{immediately} after purchase, which is far from typical as usually miners wait a long time to receive hardware. The revenue for both is \emph{after} deducting all expenses (electricity for \glspl{ASIC}, and trading fees for portfolios).

\begin{figure}
    \centering
    \begin{subfigure}[b]{0.48\linewidth}
        \centering
        \includegraphics[trim={0.0cm 0.0cm 0.0cm 0.0cm},clip,width=1\linewidth]{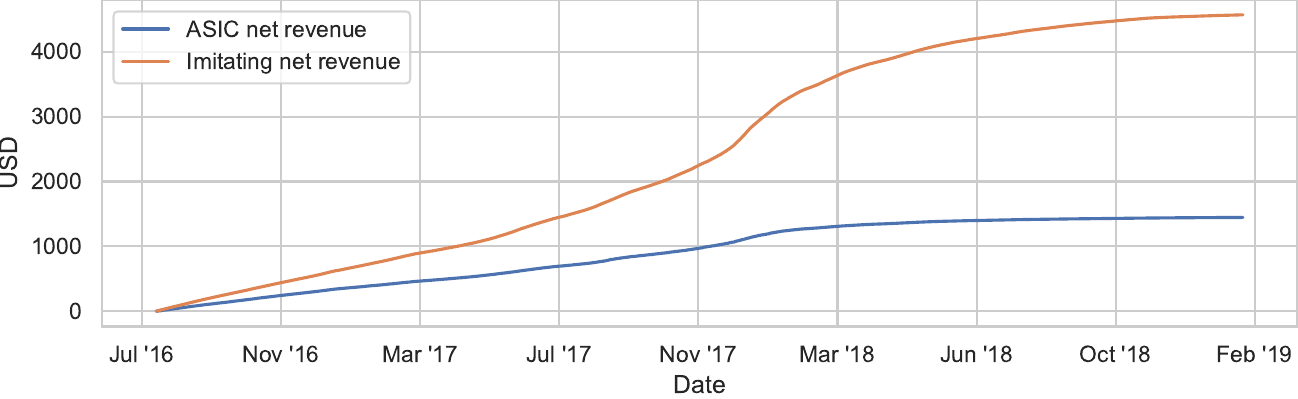}
        \caption{\gls{ASIC} and portfolio bought on July $2016$.} 
	\end{subfigure}
	\hfill
	\begin{subfigure}[b]{0.48\linewidth}
	    \centering
	    \includegraphics[trim={0.0cm 0.0cm 0.0cm 0.0cm},clip,width=1\linewidth]{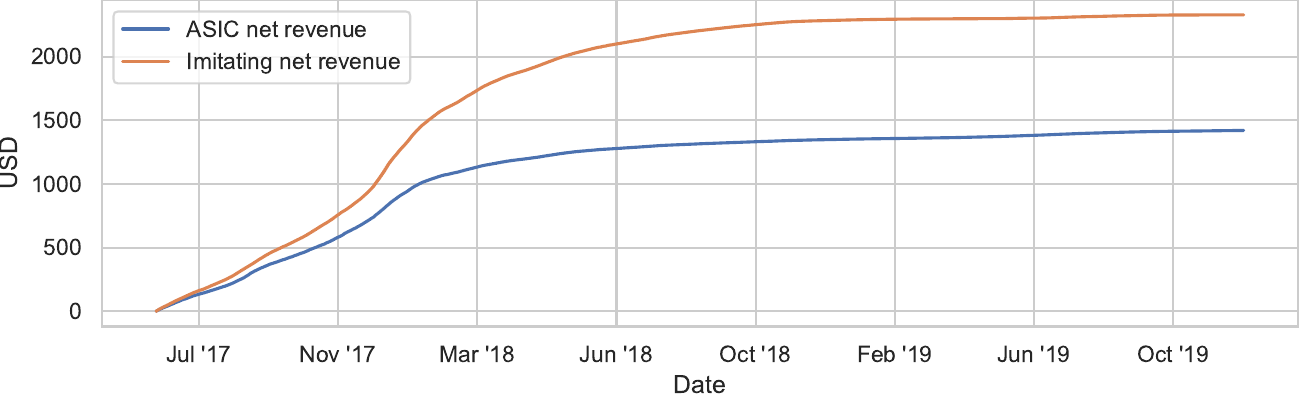}
    	\caption{\gls{ASIC} and portfolio bought on June $2017$.} 
    \end{subfigure}
    \begin{subfigure}[b]{0.48\linewidth}
        \centering
        \includegraphics[trim={0.0cm 0.0cm 0.0cm 0.0cm},clip,width=1\linewidth]{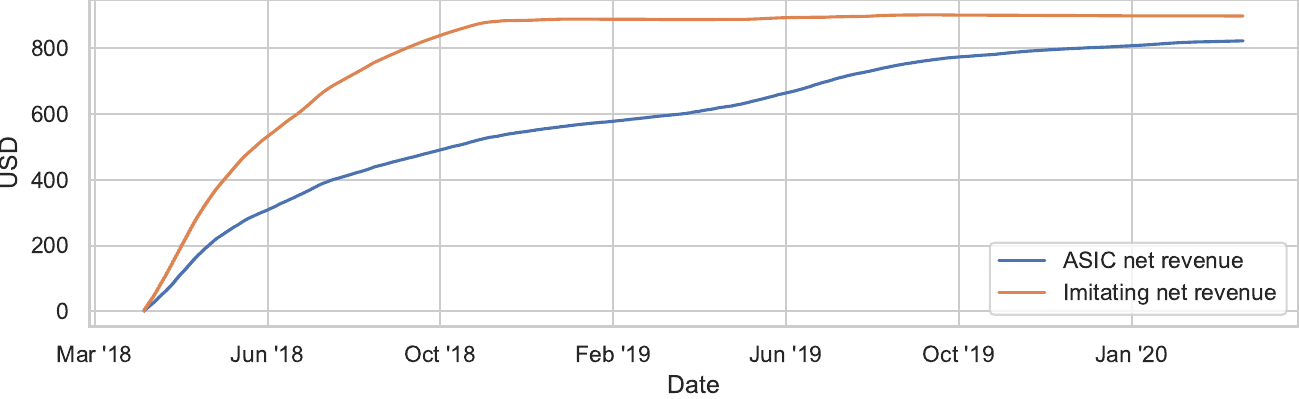}
    	\caption{\gls{ASIC} and portfolio bought on April $2018$.}
	\end{subfigure}
	\hfill
	\begin{subfigure}[b]{0.48\linewidth}
	    \centering
	    \includegraphics[trim={0.0cm 0.0cm 0.0cm 0.0cm},clip,width=1\linewidth]{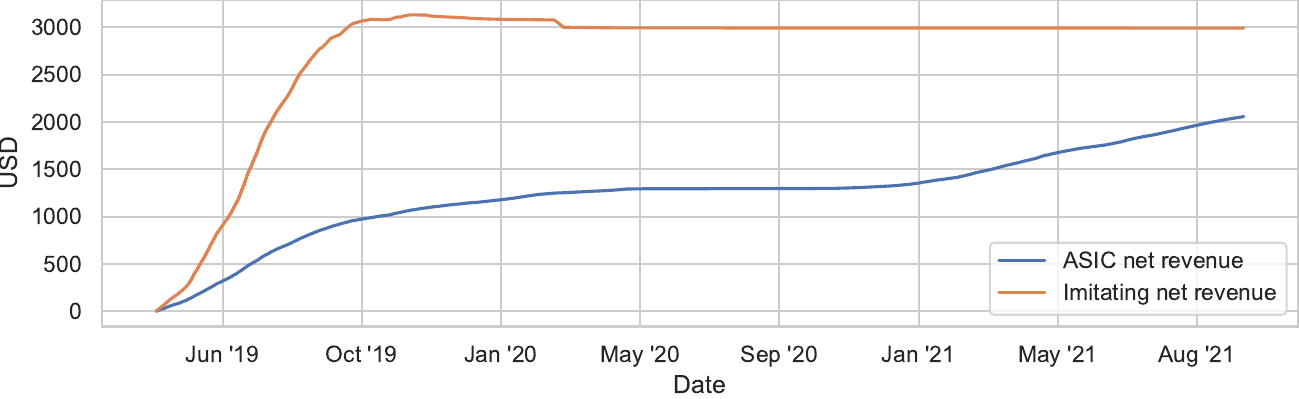}
        \caption{\gls{ASIC} and portfolio bought on May $2019$.}
    \end{subfigure}    
	\caption{Realized revenue (after expenses) of an \gls{ASIC} and its imitating portfolio, each bought for $\$1000$ at different points in time.}
	\label{fig:ImitatingComparison}
\end{figure}

Similarly, \cref{fig:ImitatingComparison} compares the realized revenue (after expenses) obtained from investing $\$1000$ in an imitating portfolio with an equal investment in real mining hardware.

Our imitating portfolio's revenue is not equal to an \gls{ASIC}'s because of a gap between the realized and projected growth rates of the network's total hash-rate.
Also, a portfolio's accuracy increases with the granularity of its time-steps, while the adjustments made at every step might increase its cost. We used $25$ steps per mining opportunity, which empirically produces accurate results.

To provide an additional angle on these results, we show in \cref{fig:AsicValueComparison} both the correct prices of mining hardware over time, and the total cost of the corresponding imitating portfolios, including the average-case fees paid for all necessary adjustments. Although an imitating portfolio is more expensive when compared to the correct price, it still costs less than the official price.

\begin{figure}
    \centering
    \begin{subfigure}[b]{0.48\linewidth}
        \centering
    	\includegraphics[trim={0cm 0.0cm 0.0cm 0.0cm},clip,width=1\linewidth]{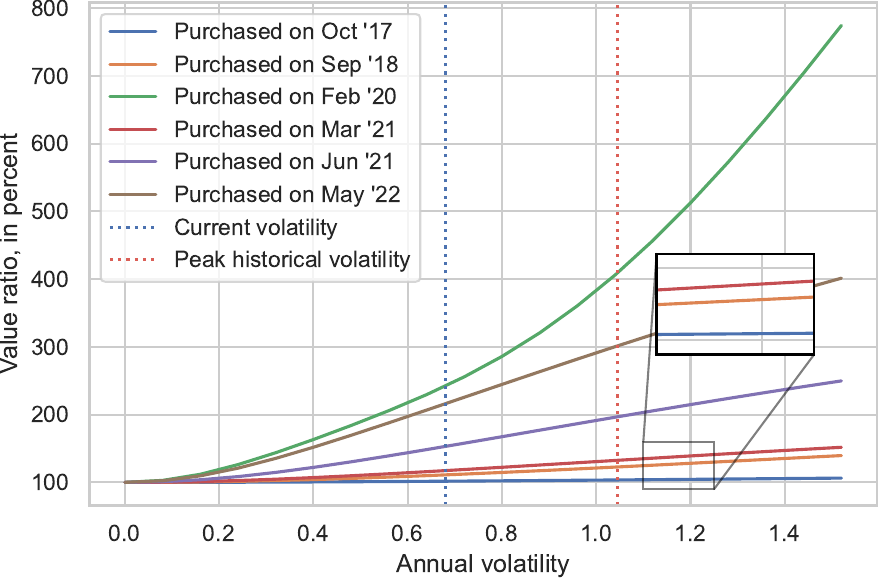}
    	\caption{Higher volatility increases \gls{ASIC} value.}
    	\label{fig:ValueAsFuncOfVolatility}
    \end{subfigure}
    \hfill
    \begin{subfigure}[b]{0.48\linewidth}
        \centering
    	\includegraphics[trim={0cm 0.0cm 0.0cm 0.0cm},clip,width=1\linewidth]{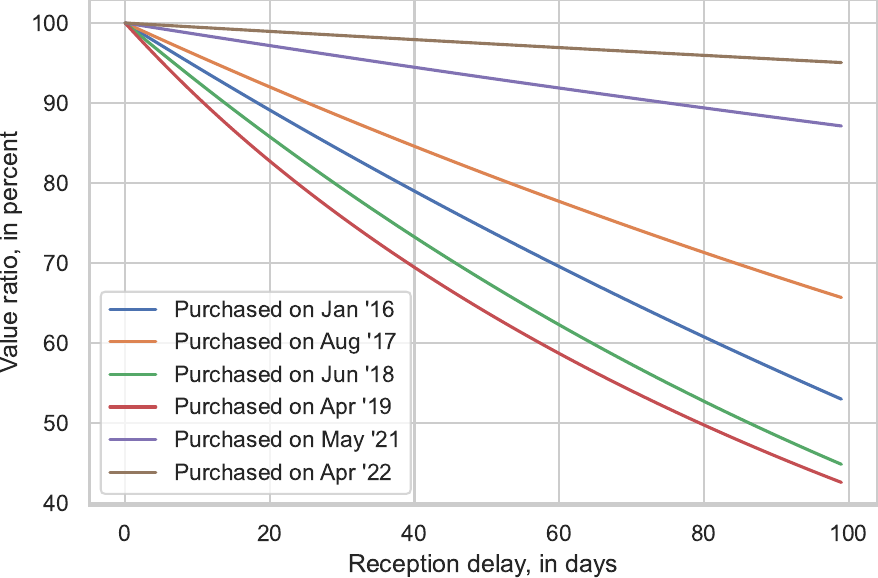}
    	\caption{\glspl{ASIC} received later in time decrease in value.}
    	\label{fig:ValueAsFuncOfDelay}
    \end{subfigure}
    \caption{The effects of volatility and delay on hardware value.}
\end{figure}

\paragraph*{Volatility Increases Value}
\cref{fig:ValueAsFuncOfVolatility} depicts our evaluation of \gls{ASIC} prices as a function of volatility, where each line represents a different purchase date. Bitcoin's annual volatility, as estimated on September $9$th, $2021$, and its peak annual volatility, which occurred in the year preceding April $29$th, $2018$, are depicted as vertical lines.

Our method gives higher prices for \glspl{ASIC} when volatility is higher. For example, an \gls{ASIC} bought on June $2021$ could cost $20\%$ more if the volatility was at its historical peak.

Of note is the increase in value for hardware bought on February $2020$. This can be explained by the crash in global hashrate experienced at the beginning of 2021 (see \cref{fig:BtcToUsdAndHash}). The combination of high volatility and low hashrate means that it is profitable to turn on hardware which might not be the most efficient or powerful (equivalently, the hardware's shutdown price is lower).

\paragraph*{Reception Delay Decreases Value}
By applying \cref{eq:ReceptionDelay} on historical data, we learn that a delay in the reception of an \gls{ASIC} can severely lower its value, with a month's delay decreasing value by $30\%$, as seen in \cref{fig:ValueAsFuncOfDelay}.

\section{Related Work}
\label{sec:RelatedWork}
To the best of our knowledge, our work is the first to evaluate the price of mining hardware, and to show that mining hardware can be imitated by purely financial means.

\paragraph*{Economic Models of Mining}
Other works have attempted to model the economics of mining without evaluating hardware prices, but most of these have not addressed the risk inherent in exchange-rate fluctuations and their affect on the economics of mining.
For example \cite{hayes2014decision,hayes2017cryptocurrency,mirkin2020bdos}, examine mining revenue in an economic setting where different cryptocurrencies co-exist.
Several papers explore single-token economic models of mining, but most focus on the willingness of new miners to enter the market based on \emph{expected} returns, and usually consider equilibria in a single shot interaction, e.g. \cite{arnosti2022bitcoin,dimitri2017bitcoin}, or works such as \cite{singh2020game}, which consider a myopic Nash equilibrium in a game model of the bitcoin market.
An equilibrium of miners in a bounded horizon setting is explored in \cite{fiat2019energy,goren2019mind}, both show that miners may gain by turning their hardware on and off repeatedly, thereby taking advantage of difficulty adjustments. 

Some tried accounting for risk, for example \cite{athey2016bitcoin}, where the price of bitcoin (but not of mining hardware) is based on user adoption and friction due to exchange-rate uncertainty, or \cite{bissias2018using} which focuses on estimating hashrate allocation between multiple tokens by using miner risk-preference to estimate their expected revenue, which our method shows can give an incorrect result.

Concurrently and independently of our work, \cite{hashimoto2019pricing} consider mining hardware as an option, but present a simpler model that lacks several factors inherent to the mining market such as: changing electricity costs, hardware decay and delivery delays. Our work also adds an empirical evaluation of the model compared to historical data and an analysis of the performance of imitating portfolios.

\paragraph*{Economic Models of Cryptocurrency Security}
An analysis of Bitcoin's security in a model where miner rewards are based on transaction fees and block-rewards are negligible is carried out in \cite{tsabary2018gap}.
An economic analysis of the security of Bitcoin is performed by \cite{budish2018economic}, arguing that when the currency is under attack, its value drops, causing mining hardware to lose value. In \cite{yaish2023uncle}, it is shown that a malicious mining strategy strictly dominates the honest one in Ethereum-like cryptocurrencies, meaning that attacking the cryptocurrency is riskless when compared to the ``honest'' mining protocol, but can earn more profits.

\paragraph*{Improving Mining Performance and Mining Pools}
Some works attempted to improve the performance of mining machines \cite{anishdev2014bitcoin,hanke2016asicboost,suresh2018optimized}, thus also increasing profits. But, these do not analyze the value of mining hardware.

A different approach is for so-called ``solo'' miners to operate as part of mining pools, which are coalitions of miners who mine together to get a steadier revenue-flow.
Indeed, most mining is performed by pools \cite{wang2020measurement}; thus, risk-aversion is believed to be widespread among miners. The economics of pools were examined by \cite{rosenfeld2011analysis,schrijvers2017incentive,salimitari2017profit}, which again neglected risk.

\section{Conclusion}
\label{sec:conclusion}
In this paper we show that widespread notions regarding \gls{ASIC} prices and their dependence on subjective measures like projected expected exchange-rates are flawed. Instead, we present a method for \emph{correctly} pricing mining hardware, and show \glspl{ASIC} can be \emph{imitated} using bonds and tokens.

Popular opinion holds that as Bitcoin becomes more widely used, its volatility will decrease. Our evaluation shows that a decrease in volatility negatively affects the value of hardware, while at the same time making imitating portfolios cheaper to maintain (smaller adjustments are needed). Combined, both negate the financial incentives put in place to encourage mining. As Bitcoin's security relies on miner participation, lower mining revenues hurt security and undermine Bitcoin's usage as a currency.

\paragraph*{Future Work}
The security risk inherent in lower volatility can be addressed by adopting random reward mechanisms to artificially increase volatility: if rewards are made to follow a random walk, the returns of miners become more volatile, thus increasing potential profits and miner participation.
To prevent miners from foreseeing future profits and stopping mining, rewards should be determined post-hoc.

We assumed that the global hash-rate is exogenous to the model, a possible extension could be to endogenize this. Miners may purchase hardware as long as it remains profitable to do so. 
Another interesting extension is to consider mining hardware capable of mining multiple currencies. 

These additions could allow using our results to estimate the global-hash rate as dependent on the reward and difficulty adjustment mechanisms of a coin and its competitors, potentially helping to design better ones that avoid pitfalls like selfish-mining and "hash-wars". 
Hash-rate could also be analyzed in relation to a coin's exchange-rate, which are correlated according to anecdotal evidence, see \cref{fig:BtcToUsdAndHash}.

\bibliography{bibliography}

\appendix

\section{Proofs}
\label[appendix]{sec:Proofs}
In this section, we prove the theorems and claims used throughout the paper.

\subsection{Pricing a Single Mining Opportunity}
\claimShortAmount*
\begin{proof}[Proof of \cref{clm:ShortAmount}]
	There are only two possible future world states: one where the coin's exchange-rate will up relative to $\finalturn-1$ and will be $\upcoinprice \coinprice{\finalturn-1}$, and the other where it will go down to $\downcoinprice \coinprice{\finalturn-1}$.
	Denote the immediate value of the mining opportunity in the up state as:
	\begin{equation} \label{eq:oppValUpImmediate}
    	\optionvalue{\finalturn, \finalturn, \upcoinprice \coinprice{\finalturn-1}} = \max \left( \frac{\hashrate \blockreward{\finalturn} \upcoinprice \coinprice{\finalturn-1}}{\globalhashrate{\finalturn} + \hashrate}  - \hashrate \efficiency \electricityprice{\finalturn}, 0 \right)
	\end{equation}
	And of the down state as:
	\begin{equation} \label{eq:oppValDownImmediate}
    	\optionvalue{\finalturn, \finalturn, \downcoinprice \coinprice{\finalturn-1}} = \max \left( \frac{\hashrate \blockreward{\finalturn} \downcoinprice \coinprice{\finalturn-1}}{\globalhashrate{\finalturn} + \hashrate}  - \hashrate \efficiency \electricityprice{\finalturn}, 0 \right)
	\end{equation}
	
	Given that $\finalturn$ is in the future, our model assumes that there is some estimation for $\globalhashrate{\finalturn}$; \cref{sec:EmpiricalEvaluation} elaborates on the way such estimates were made. Thus, the sole difficulty in evaluating $\portfoliovalue{\finalturn}$ is that although at $\finalturn-1$ we know what the value of $\coinprice{\finalturn-1}$ is, we do not yet know the realization of $\coinprice{\finalturn}$. To circumvent this, we construct the portfolio such that its value at $\finalturn$ will be the same no matter if $\coinprice{\finalturn}$ is equal to $\upcoinprice \coinprice{\finalturn-1}$ or $\downcoinprice \coinprice{\finalturn-1}$, yielding a risk-free portfolio. 
	
	The portfolio's value at the up-state is:
	\begin{equation}
	    \label{eq:PortfolioValCurrentTurnUpState}
    	\portfoliovalue{\finalturn} = \optionvalue{\finalturn, \finalturn, \upcoinprice \coinprice{\finalturn-1}} - \shortcoins{\finalturn-1} \upcoinprice \coinprice{\finalturn-1}
	\end{equation}
	And, at the down-state:
	\begin{equation}
	    \label{eq:PortfolioValCurrentTurnDownState}
    	\portfoliovalue{\finalturn} = \optionvalue{\finalturn, \finalturn, \downcoinprice \coinprice{\finalturn-1}} - \shortcoins{\finalturn-1} \downcoinprice \coinprice{\finalturn-1}
	\end{equation}
	So, we require that the following equality will hold:
	\begin{align}
	    \label{eq:PortfolioValEqual}
    	\optionvalue{\finalturn, \finalturn, \upcoinprice \coinprice{\finalturn-1}} - \shortcoins{\finalturn-1} \upcoinprice \coinprice{\finalturn-1} 
    	=
    	\optionvalue{\finalturn, \finalturn, \downcoinprice \coinprice{\finalturn-1}} - \shortcoins{\finalturn-1} \downcoinprice \coinprice{\finalturn-1}
	\end{align}
	In the above, everything but $\shortcoins{\finalturn-1}$ is known, thus it is possible to derive $\shortcoins{\finalturn-1}$ by isolating it, producing the following short amount:
	\begin{equation} \label{eq:ShortCoinsRiskFree}
    	\shortcoins{\finalturn-1} = \frac{\optionvalue{\finalturn, \finalturn, \upcoinprice \coinprice{\finalturn-1}} - \optionvalue{\finalturn, \finalturn, \downcoinprice \coinprice{\finalturn-1}}}{\coinprice{\finalturn-1} \left( \upcoinprice - \downcoinprice \right)}
	\end{equation}
	Note that there is no probability in the equation, meaning that this shorting strategy is not dependent on the probability of an upward or downward change in the coin's price. 
	
	From \cref{eq:PortfolioValCurrentTurnUpState,eq:PortfolioValCurrentTurnDownState,eq:PortfolioValEqual} we get that by doing this short, our portfolio's value at turn $\finalturn$ is:
	\begin{equation} \label{eq:PortfolioValCurrentTurn-shortCoins}
    	\portfoliovalue{\finalturn} 
    	= \optionvalue{\finalturn, \finalturn, \upcoinprice \coinprice{\finalturn-1}} - \shortcoins{\finalturn-1} \upcoinprice \coinprice{\finalturn-1}
	\end{equation}
	The equation holds in all possible world state, so the portfolio is indeed risk-free. By substituting for the short amount the following explicit form is obtained:
	\begin{align}
	    \label{eq:PortfolioValCurrentTurn-explicit}
    	\portfoliovalue{\finalturn}
    	&=
    	\frac{\optionvalue{\finalturn, \finalturn, \upcoinprice \coinprice{\finalturn-1}} - \optionvalue{\finalturn, \finalturn, \downcoinprice \coinprice{\finalturn-1}}}{\upcoinprice - \downcoinprice} \upcoinprice
    	\nonumber \\
    	&-
    	\optionvalue{\finalturn, \finalturn, \upcoinprice \coinprice{\finalturn-1}}
	\end{align}
\end{proof}

\theoremRiskFreePortfolioReturn*
\begin{proof}[Proof of \cref{thm:RiskFreePortfolioReturn}]
    \label{proof:RiskFreePortfolioReturn}
	The proof mainly relies on the no-arbitrage assumption.
	First, we define the multiplicative return of our portfolio between $\finalturn-1$ and $\finalturn$ as:
	\begin{equation} \label{eq:MultiplicativeReturn}
	    \portfolioreturn{\finalturn} \define \frac{\portfoliovalue{\finalturn}}{\portfoliovalue{\finalturn-1}}
	\end{equation}
	Thus, we want to prove that $\portfolioreturn{\finalturn} = \riskfreerate$. Assume by contradiction that $\portfolioreturn{\finalturn} \ne \riskfreerate$. We now show how to make risk-free profit in every world state by dividing to cases:
	
	\begin{enumerate} 
		\item \emph{If $\portfoliovalue{\finalturn-1} > 0$.}
		\label{case:PortfolioGreaterZero}
		
		Make a further sub-division to two sub-cases:
		
		\begin{enumerate} 
			\item \emph{If $\portfolioreturn{\finalturn} > \riskfreerate$.}
			\label{case:PortfolioGreaterZeroReturnHigher}
			
			It is possible to "make money out of nothing" by borrowing enough money at the risk-free rate to buy the portfolio at time $\finalturn-1$, and selling it after a single turn.
			
			Buying the portfolio is simply purchasing the mining opportunity and shorting the coins as specified by the portfolio, and selling it is the "reverse" - selling the opportunity and delivering the shorted asset. A reminder: shorting an asset means borrowing it and immediately selling it, thus the same asset should be returned to the loaner.
			
			Borrowing at the risk-free rate means that there is interest to be paid for the loan, but as this case assumes that the return of the portfolio is higher, a profit has been made even after taking interest into account, a contradiction to the no-arbitrage assumption.
			
			\item \emph{If $\portfolioreturn{\finalturn} < \riskfreerate$.}
			\label{case:PortfolioGreaterZeroReturnLower}
			
			Risk-less profit can be made by shorting the portfolio and investing the resulting money in a risk-free instrument at time $\finalturn-1$, and by returning the short at the next turn. 
			
			Shorting the portfolio entails shorting the mining opportunity and buying the coins, as specified by the portfolio. Returning this short is simply returning the opportunity and selling the coins.
			
			By the current case's assumption, the return on the coins and risk-free investment is large enough make a profit, even after delivering the short, and we have reached a contradiction.
		\end{enumerate}
		
		\item \emph{If $\portfoliovalue{\finalturn-1} = 0$.}
		
		$\portfolioreturn{\finalturn}$ is undefined, thus a split to different cases than before is required:
		
		\begin{enumerate} 
			\item \emph{If $\portfoliovalue{\finalturn} > 0$.}
			
			Buy the portfolio at turn $\finalturn-1$. According to the assumption of the current case, at $\finalturn-1$ the portfolio is priced at 0, meaning that shorting the required number of coins as specified in \cref{clm:ShortAmount} produces exactly enough money to buy the mining opportunity. 
			By selling the portfolio after a single turn, a risk-less profit can be made, as according to our assumptions:
			\begin{equation}
			    \portfoliovalue{\finalturn} > 0 = \portfoliovalue{\finalturn-1}
			\end{equation}
			
			\item \emph{If $\portfoliovalue{\finalturn} < 0$.}
			Short the portfolio at $\finalturn-1$ and return it after a single turn. By combining this case's assumptions:
			\begin{equation}
			    \portfoliovalue{\finalturn-1} = 0 > \portfoliovalue{\finalturn}
			\end{equation}
			After one turn the portfolio has made a loss and the short has made a risk-less profit.
			
			\item \emph{If $\portfoliovalue{\finalturn} = 0$.}
			From our assumptions we get: 
			\begin{equation}
			    \portfoliovalue{\finalturn} = 0 =\riskfreerate \portfoliovalue{\finalturn-1}
			\end{equation}
		\end{enumerate}
		
		\item \emph{If $\portfoliovalue{\finalturn-1} < 0$.}
		Proceeding as in \ref{case:PortfolioGreaterZero}:
		\begin{enumerate} 
			\item \emph{If $\portfolioreturn{\finalturn} > \riskfreerate$.}
			
			Borrow enough money at the risk-free rate to short the portfolio (this costs money in the current state). After one turn, return the short, receive $-\portfoliovalue{\finalturn}$, and pay back $-\riskfreerate \portfoliovalue{\finalturn-1}$ to repay the loan.
			
			As $\riskfreerate > 1$, we get that
			\begin{equation}
			    \portfolioreturn{\finalturn} > \riskfreerate > 1
			\end{equation}
			
			Thus, from our assumption that $\portfoliovalue{\finalturn-1} < 0$ and from the return's definition in \cref{eq:MultiplicativeReturn}:
			\begin{align}
    			\portfoliovalue{\finalturn}
    			&=
    			\portfolioreturn{\finalturn} \portfoliovalue{\finalturn-1}
    			\nonumber \\
    			&<
    			\riskfreerate \portfoliovalue{\finalturn-1}
    			\nonumber \\
    			&<
    			\portfoliovalue{\finalturn-1}
    			\nonumber \\
    			&<
    			0
			\end{align}
			We deduce that conversely: 
			\begin{align}
    			- \portfoliovalue{\finalturn}
    			&=
    			- \portfolioreturn{\finalturn} \portfoliovalue{\finalturn-1}
    			\nonumber \\
    			&>
    			- \riskfreerate \portfoliovalue{\finalturn-1} 
    			\nonumber \\
    			&>
    			- \portfoliovalue{\finalturn-1}
    			\nonumber \\
    			&>
    			0
			\end{align}
			Meaning that a risk-less profit has been made.
			
			\item \emph{If $0 \le \portfolioreturn{\finalturn} < \riskfreerate$.} \label{case:PortfolioLowerZeroReturnLower}
			
			Buy the portfolio at the first turn. As the portfolio cost is negative, buying it generates money; invest it at the risk-free rate for a single turn.
			
			At the next turn, sell the portfolio. This costs a positive amount, according to the current world state, specifically $-\portfoliovalue{\finalturn}$.
			From the assumptions and the definition of the return as given by \cref{eq:MultiplicativeReturn}:
			\begin{align}
			    \riskfreerate \portfoliovalue{\finalturn-1}
			    &<
			    \portfolioreturn{\finalturn} \portfoliovalue{\finalturn-1}
			    \nonumber \\
			    &=
			    \portfoliovalue{\finalturn}
			    \nonumber \\
			    &\le
			    0
			\end{align}
			So, we deduce:
			\begin{align}
			    -\riskfreerate \portfoliovalue{\finalturn-1}
			    &>
			    -\portfolioreturn{\finalturn} \portfoliovalue{\finalturn-1}
			    \nonumber \\
			    &=
			    -\portfoliovalue{\finalturn}
			    \nonumber \\
			    &\geq
			    0
			\end{align}
			
			$-\portfoliovalue{\finalturn}$ was lost by selling the portfolio, but the risk-free investment is worth $-\riskfreerate \portfoliovalue{\finalturn-1}$, enough to make a profit even after selling.
			
			\item \emph{If $\portfolioreturn{\finalturn} < 0$.}
			As before, by buying the portfolio at the beginning, money is earned, and it can be invested at the risk-free rate. By the next turn, the portfolio is already worth a positive amount of money, thus selling it earns even more money. So, a risk-free profit was made.
		\end{enumerate}
	\end{enumerate}
	
	All in all, if the return of the portfolio is not exactly the risk-free rate, there is an arbitrage opportunity and it is possible to make a sure profit in every world state, in contradiction to the no-arbitrage assumption; thus, the return has to equal the risk-free rate.
\end{proof}


\corollarySingleOpportunityValue*
\begin{proof}[Proof of \cref{cly:SingleOpportunityValue}] 
	According to \cref{thm:RiskFreePortfolioReturn}:
	\begin{equation}
	    \portfoliovalue{\finalturn} = \riskfreerate \portfoliovalue{\finalturn-1}
	\end{equation}
	By rearranging, we get:
	\begin{equation}
	    \portfoliovalue{\finalturn-1} = \frac{\portfoliovalue{\finalturn}} {\riskfreerate}
	\end{equation}
	Substituting by the definition of $\portfoliovalue{\finalturn-1}$ given in \cref{eq:PortfolioValPreviousTurn}:
	\begin{equation}
	    \optionvalue{\finalturn, \finalturn-1, \coinprice{\finalturn-1}} - \shortcoins{\finalturn-1} \coinprice{\finalturn-1} = \frac{\portfoliovalue{\finalturn}} {\riskfreerate} 
	\end{equation}
	We are interested in $\optionvalue{\finalturn, \finalturn-1, \coinprice{\finalturn-1}}$, so we isolate it:
	\begin{equation} \label{eq:oppValueOneTurnBefore}
	    \optionvalue{\finalturn, \finalturn-1, \coinprice{\finalturn-1}} = \shortcoins{\finalturn-1} \coinprice{\finalturn-1} + \frac{\portfoliovalue{\finalturn}}{\riskfreerate}
	\end{equation}
	By using \cref{eq:PortfolioValCurrentTurn-shortCoins} to substitute for $\portfoliovalue{\finalturn}$:
	\begin{align}
    	\optionvalue{\finalturn, \finalturn-1, \coinprice{\finalturn-1}}
    	&=
    	\frac{1}{\riskfreerate} \left( \optionvalue{\finalturn, \finalturn, \upcoinprice \coinprice{\finalturn-1}} - \shortcoins{\finalturn-1} \upcoinprice \coinprice{\finalturn-1} \right)
    	\nonumber \\
    	&+
    	\shortcoins{\finalturn-1} \coinprice{\finalturn-1}
	\end{align}
	Slightly rearranging:
	\begin{align}
    	\optionvalue{\finalturn, \finalturn-1, \coinprice{\finalturn-1}}
    	&=
    	\shortcoins{\finalturn-1} \coinprice{\finalturn-1} \left( 1 - \frac{\upcoinprice}{\riskfreerate} \right)
    	\nonumber \\
    	&+
    	\frac{\optionvalue{\finalturn, \finalturn, \upcoinprice \coinprice{\finalturn-1}}}{\riskfreerate}
	\end{align}
	Finally, substituting for $\shortcoins{\finalturn-1}$ as given in \cref{clm:ShortAmount}, an explicit form is reached:
	\begin{align}
	    \label{eq:oppValueOneTurnBefore-explicit}
    	\optionvalue{\finalturn, \finalturn-1, \coinprice{\finalturn-1}}
    	&=
    	\frac{\optionvalue{\finalturn, \finalturn, \upcoinprice \coinprice{\finalturn-1}} - \optionvalue{\finalturn, \finalturn, \downcoinprice \coinprice{\finalturn-1}}}{\upcoinprice - \downcoinprice} \left( 1 - \frac{\upcoinprice}{\riskfreerate} \right)  
    	\nonumber \\
    	&+
    	\frac{\optionvalue{\finalturn, \finalturn, \upcoinprice \coinprice{\finalturn-1}}}{\riskfreerate}
	\end{align}
	
	Note that all factors are known and can be calculated at time $\finalturn-1$. Specifically, $\optionvalue{\finalturn, \finalturn, \upcoinprice \coinprice{\finalturn-1}}$ and $\optionvalue{\finalturn, \finalturn, \downcoinprice \coinprice{\finalturn-1}}$ can be obtained by substituting for the correct exchange-rate in \cref{eq:oppValImmediate}. 
\end{proof}

\subsection{Pricing Relative to an Arbitrary Time}

\claimRiskFreePortfolioArbitraryTurn*
\begin{proof}[Proof of \cref{clm:RiskFreePortfolioArbitraryTurn}] 
	At turn $\turnReg$ it seems the uncertainty regarding the coin's exchange rate at turn $\finalturn$ is larger because there are $\finalturn - \turnReg + 1$ possible "final" future values instead of only $2$, as shown in \cref{fig:PriceTree} for the case where $\finalturn = 2$.
	
	But, luckily, we are given $\optionvalue{\finalturn, \turnReg+1, \upcoinprice \coinprice{\turnReg}}$ and $\optionvalue{\finalturn, \turnReg+1, \downcoinprice \coinprice{\turnReg}}$. We use both values to construct a risk-free portfolio such that its value at $\turnReg+1$ will be the same no matter if the exchange-rate will go up or down. Similarly to \cref{clm:ShortAmount}, it will hold the $\finalturn$-th opportunity, and a short on $\shortcoins{\turnReg}$ coins.
	
	At turn $\turnReg+1$ the portfolio's value is defined by:
	\begin{equation}
	    \label{eq:PortfolioValCurrentTurnArbitraryTime}
	    \portfoliovalue{\turnReg} = \optionvalue{\finalturn, \turnReg, \coinprice{\turnReg}} - \shortcoins{\turnReg} \coinprice{\turnReg}
	\end{equation}
	And at $\turnReg+1$ it is:
	\begin{equation} \label{eq:PortfolioValNextTurnArbitraryTime}
	    \portfoliovalue{\turnReg+1} = \optionvalue{\finalturn, \turnReg+1, \coinprice{\turnReg+1}} - \shortcoins{\turnReg} \coinprice{\turnReg+1}
	\end{equation}
	If the coin's exchange-rate has moved upwards between $\turnReg, \turnReg+1$, the portfolio will be worth:
	\begin{equation}
	    \portfoliovalue{\turnReg+1} = \optionvalue{\finalturn, \turnReg+1, \upcoinprice \coinprice{\turnReg}} - \shortcoins{\turnReg} \upcoinprice \coinprice{\turnReg}
	\end{equation}
	Similarly for the down-state:
	\begin{equation}
	    \portfoliovalue{\turnReg+1} = \optionvalue{\finalturn, \turnReg+1, \downcoinprice \coinprice{\turnReg}} - \shortcoins{\turnReg} \downcoinprice \coinprice{\turnReg}
	\end{equation}
	So, to make it risk-free the following property should hold:
	\begin{equation} 
	    \optionvalue{\finalturn, \turnReg+1, \upcoinprice \coinprice{\turnReg}} - \shortcoins{\turnReg} \upcoinprice \coinprice{\turnReg} = \optionvalue{\finalturn, \turnReg+1, \downcoinprice \coinprice{\turnReg}} - \shortcoins{\turnReg} \downcoinprice \coinprice{\turnReg}
	\end{equation}
	Solving for $\shortcoins{\turnReg}$ gives the following short:
	\begin{equation}
	    \label{eq:ShortCoinsRiskFreeArbitraryTime}
	    \shortcoins{\turnReg} = \frac{\optionvalue{\finalturn, \turnReg+1, \upcoinprice \coinprice{\turnReg}} - \optionvalue{\finalturn, \turnReg+1, \downcoinprice \coinprice{\turnReg}}}{\coinprice{\turnReg} \left( \upcoinprice - \downcoinprice \right)}
	\end{equation}
	Exactly the same as in the proof for \cref{thm:RiskFreePortfolioReturn}, the return of the portfolio at turn $\turnReg+1$ is equal to $\riskfreerate$:
	\begin{equation} \label{eq:RiskFreePortfolioReturnArbitraryTime}
	    \portfoliovalue{\turnReg+1} = \riskfreerate \portfoliovalue{\turnReg}
	\end{equation}
	Thus, by employing similar reasoning to \cref{cly:SingleOpportunityValue} it is possible to derive the result:
	\begin{align}
	    \label{eq:oppValueArbitraryTurnBefore-explicit}
    	\optionvalue{\finalturn, \turnReg, \coinprice{\turnReg}}
    	&=
    	\frac{\optionvalue{\finalturn, \turnReg+1, \upcoinprice \coinprice{\turnReg}} - \optionvalue{\finalturn, \turnReg+1, \downcoinprice \coinprice{\turnReg}}}{\upcoinprice - \downcoinprice} \left( 1 - \frac{\upcoinprice}{\riskfreerate} \right) 
    	\nonumber \\
    	&+
    	\frac{\optionvalue{\finalturn, \turnReg+1, \upcoinprice \coinprice{\turnReg}}}{\riskfreerate}
	\end{align}
\end{proof}

\theoremOppValueEquation*
\begin{proof}[Proof for \cref{thm:oppValueEquation}]
	Let $\firstturn < \finalturn$. We start by applying \cref{clm:RiskFreePortfolioArbitraryTurn} on $\optionvalue{\finalturn, \firstturn, \coinprice{\firstturn}}$:
	\begin{align}
    	\optionvalue{\finalturn, \firstturn, \coinprice{\firstturn}}
    	&=
    	\frac{\optionvalue{\finalturn, \firstturn+1, \upcoinprice \coinprice{\firstturn}} - \optionvalue{\finalturn, \firstturn+1, \downcoinprice \coinprice{\firstturn}}}{\upcoinprice - \downcoinprice} \left( 1 - \frac{\upcoinprice}{\riskfreerate} \right) 
    	\nonumber \\
    	&+
    	\frac{\optionvalue{\finalturn, \firstturn+1, \upcoinprice \coinprice{\firstturn}}}{\riskfreerate}
	\end{align}
	Note that $\optionvalue{\finalturn, \firstturn+1, \upcoinprice \coinprice{\firstturn}}$ appears in multiple places, by gathering all occurrences we get:
	\begin{align}
    	\optionvalue{\finalturn, \turnReg, \coinprice{\turnReg}}
    	&=
    	\left( \frac{1 - \frac{\upcoinprice} {\riskfreerate}}{\upcoinprice - \downcoinprice} + \frac{1}{\riskfreerate} \right) 
    	\optionvalue{\finalturn, \turnReg+1, \upcoinprice \coinprice{\turnReg}}
    	\nonumber \\
    	&-
    	\left( \frac{1 - \frac{\upcoinprice} {\riskfreerate}}{\upcoinprice - \downcoinprice} \right)
    	\optionvalue{\finalturn, \turnReg+1, \downcoinprice \coinprice{\turnReg}} 
	\end{align}
	Denote $\downcoefficient = \frac{1 - \frac{\upcoinprice} {\riskfreerate}}{\upcoinprice - \downcoinprice}$, and $ \upcoefficient = \downcoefficient + \frac{1}{\riskfreerate} $, then:
	\begin{equation}
    	\optionvalue{\finalturn, \firstturn, \coinprice{\firstturn}} = 
    	\upcoefficient \optionvalue{\finalturn, \firstturn+1, \upcoinprice \coinprice{\firstturn}}
    	- \downcoefficient \optionvalue{\finalturn, \firstturn+1, \downcoinprice \coinprice{\firstturn}}
	\end{equation}
	The opportunity's value is now represented as a recursive formula. Let us repeat the previous steps recursively on $\optionvalue{\finalturn, \firstturn+1, \upcoinprice \coinprice{\firstturn}}$ and $\optionvalue{\finalturn, \firstturn+1, \downcoinprice \coinprice{\firstturn}}$:
	\begin{align}
    	\optionvalue{\finalturn, \firstturn, \coinprice{\firstturn}}
    	&=
    	\upcoefficient \cdot \Big( \upcoefficient \cdot \optionvalue{\finalturn, \firstturn+2, \upcoinprice ^ 2 \coinprice{\firstturn}}
    	\nonumber \\
    	&-
    	\downcoefficient \cdot \optionvalue{\finalturn, \firstturn+2, \downcoinprice \upcoinprice \coinprice{\firstturn}} \Big)
    	\nonumber \\
    	&-
    	\downcoefficient \cdot \Big( \upcoefficient \cdot \optionvalue{\finalturn, \firstturn+2, \upcoinprice \downcoinprice \coinprice{\firstturn}}
    	\nonumber \\
    	&-
    	\downcoefficient \cdot \optionvalue{\finalturn, \firstturn+2, \downcoinprice ^ 2  \coinprice{\firstturn}} \Big)
	\end{align}
	As $\optionvalue{\finalturn, \firstturn+2, \downcoinprice \upcoinprice \coinprice{\firstturn}}$ and $\optionvalue{\finalturn, \firstturn+2, \upcoinprice \downcoinprice \coinprice{\firstturn}}$ are equal:
	\begin{align}
    	\optionvalue{\finalturn, \firstturn, \coinprice{\firstturn}}
    	&=
    	\upcoefficient^2 \cdot \optionvalue{\finalturn, \firstturn+2, \upcoinprice ^ 2 \coinprice{\firstturn}}
    	\nonumber \\
    	&-
    	2 \cdot \upcoefficient \cdot \downcoefficient \optionvalue{\finalturn, \firstturn+2,  \upcoinprice \downcoinprice \coinprice{\firstturn}}
    	\nonumber \\
    	&+
    	\downcoefficient^2 \cdot \optionvalue{\finalturn, \firstturn+2, \downcoinprice ^ 2  \coinprice{\firstturn}}
	\end{align}
	We inductively continue with the recursion until reaching the exercise time of the opportunity, resulting in:
	\begin{align}
    	\optionvalue{\finalturn, \firstturn, \coinprice{\firstturn}}
    	=
    	\sum_{\turnReg=0}^{\finalturn - \firstturn}  
    	\binom{\finalturn - \firstturn}{\turnReg} \upcoefficient ^ \turnReg \left( - \downcoefficient \right)^{\finalturn - \firstturn - \turnReg}
    	\optionvalue{\finalturn, \finalturn, \upcoinprice^{\turnReg} \downcoinprice^{\finalturn-\firstturn-\turnReg} \coinprice{\firstturn}} 
	\end{align}
	Slightly rearranging:
	\begin{equation}
	    \optionvalue{\finalturn, \firstturn, \coinprice{\firstturn}} = 
	    \sum_{\turnReg=0}^{\finalturn - \firstturn}  
    	\frac{\binom{\finalturn - \firstturn}{\turnReg} \upcoefficient ^ \turnReg}
    	{\left( - \downcoefficient \right) ^{\firstturn+\turnReg-\finalturn}}
    	\optionvalue{\finalturn, \finalturn, \upcoinprice^{\turnReg} \downcoinprice^{\finalturn-\firstturn-\turnReg} \coinprice{\firstturn}} 
	\end{equation}
	Note that the sum might go over states where the opportunity's value is equal to zero, which is unnecessary and can be avoided by starting the summation only from $\turnReg$ where:
	\begin{equation} 
	    \optionvalue{\finalturn, \finalturn, \upcoinprice^{\turnReg} \downcoinprice^{\finalturn-\firstturn-\turnReg} \coinprice{\firstturn}} > 0
	\end{equation}
	By the definition given in \cref{eq:oppValImmediate}, this is the same as requiring:
	\begin{equation}
    	\max \left( 
    	\frac{ \hashrate \blockreward{\finalturn} \upcoinprice ^{\turnReg} \downcoinprice^{\finalturn - \firstturn - \turnReg} \coinprice{\firstturn}}{\globalhashrate{\finalturn} +\hashrate} - \hashrate \efficiency \electricityprice{\finalturn} 
    	, 0
    	\right)
    	> 0
	\end{equation}
	As the opportunity's value is strictly greater than $0$, the $\max$ can be dropped, resulting in:
	\begin{equation}
	    \frac{\blockreward{\finalturn} \downcoinprice^{\finalturn - \firstturn} \coinprice{\firstturn}  }{\globalhashrate{\finalturn} +\hashrate} 
	    \left( \frac{\upcoinprice}{\downcoinprice} \right)^{\turnReg} 
	    >
	    \efficiency \electricityprice{\finalturn} 
	\end{equation}
	By isolating $\turnReg$ we can find the minimal turn where this condition is held.
    First, isolate $\frac{\upcoinprice}{\downcoinprice}$:
	\begin{equation}
    	\left( \frac{\upcoinprice}{\downcoinprice} \right)^{\turnReg} 
    	>
    	\frac{\efficiency \electricityprice{\finalturn}}
    	{\left( \frac{\blockreward{\finalturn} \downcoinprice^{\finalturn - \firstturn} \coinprice{\firstturn} }{\globalhashrate{\finalturn} + \hashrate} \right)} 
    	=
    	\left( \frac{\blockreward{\finalturn} \downcoinprice^{\finalturn - \firstturn} \coinprice{\firstturn} }{\globalhashrate{\finalturn} + \hashrate} \right)^{-1} 
    	\cdot
    	\efficiency \electricityprice{\finalturn}
	\end{equation}
	Now, take the logarithm of both sides:
	\begin{equation}
    	\turnReg \cdot \log \left( \frac{\upcoinprice}{\downcoinprice} \right)
    	>
    	\log \left( 
    	\left( \frac{\blockreward{\finalturn} \downcoinprice^{\finalturn - \firstturn} \coinprice{\firstturn} }{\globalhashrate{\finalturn} + \hashrate} \right)^{-1} 
    	\efficiency \electricityprice{\finalturn}
    	\right)
	\end{equation}
	Finally, we isolate $\turnReg$:
	\begin{equation}
    	\turnReg 
    	>
    	\frac
    	{ \log \left( \left( \frac{\blockreward{\finalturn} \downcoinprice^{\finalturn - \firstturn} \coinprice{\firstturn} }{\globalhashrate{\finalturn} + \hashrate} \right)^{-1} 
    		\efficiency \electricityprice{\finalturn} \right)}
    	{\log \left( \frac{\upcoinprice}{\downcoinprice} \right)}
	\end{equation}
	So, the minimal turn for which the opportunity's value is greater than $0$ is:
	\begin{equation}
	\turnReg_0 \define \ceil*{
		\frac
		{\log \left( 
			\left( \frac{\blockreward{\finalturn}  \downcoinprice^{\finalturn - \firstturn} \coinprice{\firstturn} }{\globalhashrate{\finalturn} +\hashrate} \right) ^{-1}
			\efficiency \electricityprice{\finalturn} 
			\right)}
		{\log \left( \frac{\upcoinprice}{\downcoinprice} \right)}
	}
	\end{equation}
	Starting the summation from $\turnReg_0$ gives the following equation:
	\begin{equation} \label{eq:oppValueEquation}
    	\optionvalue{\finalturn, \firstturn, \coinprice{\firstturn}}  
    	= 
    	\sum_{\turnReg=\turnReg_0}^{\finalturn - \firstturn}  
    	\frac{\binom{\finalturn - \firstturn}{\turnReg} \upcoefficient ^ \turnReg}
    	{\left( - \downcoefficient \right) ^{\firstturn+\turnReg-\finalturn}}
    	\optionvalue{\finalturn, \finalturn, \upcoinprice^{\turnReg}  \downcoinprice^{\finalturn-\firstturn-\turnReg} \coinprice{\firstturn}} 
	\end{equation}
	As noted before, thanks to summing only strictly positive values it is possible to drop the $\max$, resulting in the following equation:
	\begin{align}
    	\optionvalue{\finalturn, \firstturn, \coinprice{\firstturn}}
    	=
    	\sum_{\turnReg=\turnReg_0}^{\finalturn-\firstturn}   
    	\frac{\binom{\finalturn-\firstturn}{\turnReg} \left( \upcoefficient \right) ^ \turnReg \hashrate} 
    	{\left( - \downcoefficient \right) ^{\firstturn+\turnReg-\finalturn}} 
    	\left( 
        	\frac{\blockreward{\finalturn} \coinprice{\firstturn} \downcoinprice^{\finalturn - \firstturn} }{\globalhashrate{\finalturn} +\hashrate} 
        	\left( \frac{\upcoinprice}{\downcoinprice} \right)^{\turnReg} - \efficiency \electricityprice{\finalturn} 
    	\right) 
	\end{align}
\end{proof}

\subsection{Imitating Portfolio}

\theoremImitatingPortfolio*
\begin{proof}
The proof proceeds via a series of claims.

\claimLongAmount*
\begin{proof}[Proof of \cref{clm:LongAmount}]
	The proof is similar to that of \cref{clm:ShortAmount}. We want the portfolio to be worth the same as the underlying asset in the next turn, no matter the realization of the exchange-rate.
	
	If the exchange-rate has went up, the portfolio's value is:
	\begin{equation}
    	\imitatingportfoliovalue{\turnReg+1} = \riskfreerate \bonds{\turnReg} + \longcoins{\turnReg} \upcoinprice \coinprice{\turnReg}
	\end{equation}
	If it went down, the value is:
	\begin{equation}
	    \imitatingportfoliovalue{\turnReg+1} = \riskfreerate \bonds{\turnReg} + \longcoins{\turnReg} \downcoinprice \coinprice{\turnReg}
	\end{equation}
	So, to find the correct values for $\bonds{\turnReg}, \longcoins{\turnReg}$ we need to solve the following system of linear equations:
	\begin{align}
	    \upcoinprice \coinprice{\turnReg} \longcoins{\turnReg} + \riskfreerate \bonds{\turnReg} &= \optionvalue{\finalturn, \turnReg+1, \upcoinprice \coinprice{\turnReg}} \\
	    \downcoinprice \coinprice{\turnReg} \longcoins{\turnReg} + \riskfreerate \bonds{\turnReg} &= \optionvalue{\finalturn, \turnReg+1, \downcoinprice \coinprice{\turnReg}}
	\end{align}
	The only solution is:
	\begin{align}
	    \label{eq:LongCoinsImitating}
    	\longcoins{\turnReg}
    	&=
    	\frac{\optionvalue{\finalturn, \turnReg+1, \upcoinprice \coinprice{\turnReg}} - \optionvalue{\finalturn, \turnReg+1, \downcoinprice \coinprice{\turnReg}}}{\coinprice{\turnReg} \left( \upcoinprice - \downcoinprice \right)}
	    \\
	    \label{eq:BondsImitating}
    	\bonds{\turnReg}
    	&=
    	\frac{\upcoinprice \optionvalue{\finalturn, \turnReg+1, \downcoinprice \coinprice{\turnReg}} - \downcoinprice \optionvalue{\finalturn, \turnReg+1, \upcoinprice \coinprice{\turnReg}}}{\riskfreerate \left( \upcoinprice - \downcoinprice \right)}
	\end{align}
\end{proof}

\claimImitatingPortfolio*
\begin{proof}[Proof of \cref{clm:ImitatingPortfolio}]
	According to \cref{clm:LongAmount} and the definition given in \cref{eq:ImitatingPortfolioValNextTurn}, the value of the portfolio at time $\turnReg+1$ is:
	\begin{equation} \label{eq:ImitatingPortfolioNextTurnEqualToOpp}
    	\imitatingportfoliovalue{\turnReg+1} = \riskfreerate \bonds{\turnReg} + \longcoins{\turnReg} \coinprice{\turnReg+1} = \optionvalue{\finalturn, \turnReg+1, \coinprice{\turnReg+1}}
	\end{equation}	
	Recall that the risk-free portfolio constructed in \cref{clm:RiskFreePortfolioArbitraryTurn} has the following value at $\turnReg+1$:
	\begin{equation}
    	\portfoliovalue{\turnReg+1} = \optionvalue{\finalturn, \turnReg+1, \coinprice{\turnReg+1}} - \shortcoins{\turnReg} \coinprice{\turnReg+1}
	\end{equation}	
	By isolating the opportunity's value we get:
	\begin{equation}
    	\optionvalue{\finalturn, \turnReg+1, \coinprice{\turnReg+1}} = \portfoliovalue{\turnReg+1} + \shortcoins{\turnReg} \coinprice{\turnReg+1}
	\end{equation}
	Thus, by substituting the above in \cref{eq:ImitatingPortfolioNextTurnEqualToOpp}:
	\begin{equation} \label{eq:ImitatingEqualsRiskFreePlusCoinsNextTurn}
	    \imitatingportfoliovalue{\turnReg+1} = \riskfreerate \bonds{\turnReg} + \longcoins{\turnReg} \coinprice{\turnReg+1} = \portfoliovalue{\turnReg+1} + \shortcoins{\turnReg} \coinprice{\turnReg+1}
	\end{equation}
	Note that the amount of coins in both portfolios of \cref{clm:LongAmount,clm:RiskFreePortfolioArbitraryTurn} is identical:
	\begin{equation} \label{eq:ShortCoins-longCoinsEqual}
	    \longcoins{\turnReg} = \frac{\optionvalue{\finalturn, \turnReg+1, \upcoinprice \coinprice{\turnReg}} - \optionvalue{\finalturn, \turnReg+1, \downcoinprice \coinprice{\turnReg}}}{\coinprice{\turnReg} \left( \upcoinprice - \downcoinprice \right)} = \shortcoins{\turnReg}
	\end{equation}
	So, both can be eliminated from \cref{eq:ImitatingEqualsRiskFreePlusCoinsNextTurn}, resulting in:
	\begin{equation}
	    \riskfreerate \bonds{\turnReg} = \portfoliovalue{\turnReg+1}
	\end{equation}
	As \cref{eq:RiskFreePortfolioReturnArbitraryTime} shows, the return of risk-free portfolio is equal to the risk-free rate:
	\begin{equation}
	    \riskfreerate \bonds{\turnReg} = \riskfreerate \portfoliovalue{\turnReg}
	\end{equation}
	From the assumption that $\riskfreerate \ne 0$, it is possible to divide by it:
	\begin{equation}
	    \bonds{\turnReg} = \portfoliovalue{\turnReg}
	\end{equation}
	This equality can be used to replace $\bonds{\turnReg}$ in \cref{eq:ImitatingportfolioValCurrentTurn}, giving:
	\begin{equation}
    	\imitatingportfoliovalue{\turnReg} = \portfoliovalue{\turnReg} + \longcoins{\turnReg} \coinprice{\turnReg}
	\end{equation}
	Substituting for $\portfoliovalue{\turnReg}$ by using \cref{eq:PortfolioValCurrentTurnArbitraryTime} gives us:
	\begin{equation}
	    \imitatingportfoliovalue{\turnReg} = \optionvalue{\finalturn, \turnReg, \coinprice{\turnReg}} - \shortcoins{\turnReg} \coinprice{\turnReg} + \longcoins{\turnReg} \coinprice{\turnReg}
	\end{equation}
	From \cref{eq:ShortCoins-longCoinsEqual}, we get:
	\begin{equation}
	    \longcoins{\turnReg} \coinprice{\turnReg} - \shortcoins{\turnReg} \coinprice{\turnReg} = 0
	\end{equation}
	Finally, we deduce:
	\begin{equation}
    	\imitatingportfoliovalue{\turnReg} = \optionvalue{\finalturn, \turnReg, \coinprice{\turnReg}}
	\end{equation}
\end{proof}

\corollaryImitating*

Now, by applying \cref{cly:Imitating} at each point in time and dynamically moving backwards until reaching $\turnReg$, we obtain the imitating portfolio for $\turnReg$ and all possible adjustments which may be required to maintain it until $\finalturn$. At this final step, all tokens and bonds contained in the portfolio should be sold. By \cref{clm:ImitatingPortfolio}, the profits made by selling the portfolio are equal to the value of the imitated mining opportunity.
\end{proof}

\section{Glossary}
\label[appendix]{sec:glossary}
Following is a list of important notations used in the paper.
\setglossarystyle{alttree}
\glssetwidest{AAAA}
\printnoidxglossary[type={symbols}]
\printnoidxglossary[type={\acronymtype}]

\end{document}